\documentclass[aps,reprint,prx,showpacs]{revtex4-2}
\usepackage{graphicx}
\usepackage{color}
\usepackage{amsmath}
\usepackage{amssymb}
\usepackage{subcaption}
\usepackage{tabularray}
\usepackage{hyperref}

\newcommand{\Min}{\textsc{Min}}
\newcommand{\Max}{\textsc{Max}}
\newcommand{\Sp}{\textsc{Sp}}

\newcommand{\inb}[3]{{\left#1 #2 \right#3}}
\newcommand{\reff}[1]{{Fig.\ref{fig:#1}}}
\newcommand{\refe}[1]{{Eq.\ref{eq:#1}}}
\newcommand{\reft}[1]{{Tab.\ref{tab:#1}}}
\newcommand{\refs}[1]{{Sec.\ref{sec:#1}}}
\newcommand{\squad}{\ \ }

\usepackage{tikz}
\newcommand{\downrightarrow}{%
  \tikz[x=1ex, y=1ex, baseline=-1.5ex]{
    \draw[-stealth, line width=0.12ex] (0,0) -- (4,-2);
  }%
}

\begin{document}

\title{Statistical Properties of Quadrangular Surfaces}
\author{Hrant Topchyan$^{1}$, Rudik Badalyan$^{2}$, Ara Sedrakyan$^{1}$}
\affiliation{
$^1$ A. Alikhanyan National Science Laboratory, Br. Alikhanian 2, Yerevan 0036, Armenia\\
$^2$ Department of Physics, University of Massachusetts, Amherst, Massachusetts 01003, USA}

\begin{abstract}
    We investigate the statistical properties of random
    quadrangular surfaces generated by different randomization procedures:
    the Gruzberg-Kl\"umper-Nuding-Sedrakyan (GKNS) construction,
    and two newly introduced generalizations of dynamical triangulations (DT),
    dynamical (DQ) and general quadrangulations (GQ).
    We formulate these surfaces within a unified graph-theoretic
    framework and establish the relationships between
    the elementary operations defining the different ensembles.
    For GKNS surfaces, we demonstrate that the construction is
    equivalent to two mutually constrained percolation processes
    and determine the associated critical point and critical exponents,
    revealing deviations from ordinary percolation.
    For DQ and GQ, we analyze the underlying Markov chains
    and determine the scaling of mixing and relaxation times.
    We further analyze all three ensembles through their
    degree distributions, degree correlations, distance statistics,
    and Hausdorff dimensions.
    While DQ exhibits exponentially decaying degree distributions
    and geometric properties similar to DT,
    GKNS and GQ display broad, scale-free degree distributions.
    Moreover,
    GKNS surfaces possess an asymptotic Hausdorff dimension $\Delta_H\approx 2$,
    whereas DQ and GQ approach $\Delta_H\approx 4$, similarly to DT.
    This indicates that DQ is compatible with the
    universal behavior of DT, while GKNS and GQ define
    distinct classes of random geometry,
    implying a different underlying measure in the space of random surfaces
    and a possible geometric framework for a new class of noncritical string theories.
\end{abstract}

\maketitle

\section{Introduction}
Discrete random surfaces became an important topic in theoretical physics in the 1980s
primarily because of their role in concretely defining and studying non-critical string theories,
i.e. string theories formulated in spacetime dimensions other than the ``critical" one
(26 for bosonic strings, 10 for superstrings) \cite{Ambjorn-1985, David-1985, Kazakov-1985}.
Non-critical string physics \cite{Polyakov-1981,Polyakov-book} is based on the idea that real,
low-dimensional gauge theories with various gauge groups can be reduced to a theory of quantum fluctuations of two-dimensional surfaces embedded in a higher-dimensional space, on which the matter degrees of freedom reside, together with the transverse degrees of freedom of the gauge fields.

The theory of non-critical strings was intensively developed toward the end of the last century. However, developments in the continuum formulation, as well as in the matrix model approach based on triangular approximations of two-dimensional surfaces, led to the result that the dynamics of these surfaces can be consistently quantized only if the central charge of the matter fields residing on them is less than or equal to one. Both approaches, the continuum limit \cite{KPZ-1988,Distler-1989,David-1988} and the study of statistical physics models on triangular lattices with central charge less than one \cite{Ambjorn-book,Kazakov-1985}, lead to fully consistent results. This demonstrates an important point, namely that the measure of two-dimensional random surfaces in the continuum limit coincides with that defined for triangular surfaces in statistical physics, with the continuum theory emerging as the scaling limit of triangular random lattices.
In this approach, random triangular surface is equivalent to surface with gravitational metric and
the summation over surfaces is  equivalent to integration over the metric as a quantum field \cite{Polyakov-1981,Polyakov-book}.

Random triangulated surfaces are of interest well beyond string theory.
They provide a minimal model for disordered geometry in a range of
condensed-matter systems \cite{Duplantier-1988, Kostov-1989, Duplantier-1989},
while also serving as the standard discrete representation of surfaces in
computational geometry, computer graphics and finite-element methods \cite{CG-tris}.
These complementary motivations have led to a vast literature
on the statistical mechanics, geometry, combinatorics and
algorithms of random triangulations \cite{Ambjorn-book,Janke-2004}.

However, the triangulation of random surfaces is not the only way to discretize a continuum surface immersed in 
$d$-dimensional Euclidean space. Another possibility is the quadrangulation of a surface.
They are emerging in various keystone  problems of theoretical physics.
The first one is the three-dimensional gauge Ising model formulated by Wegner\cite{Wegner-1971, Wilson-1974}.
Another noteworthy application of quadrangular random surfaces 
is the introduction of the GKNS construction
in the study of the integer quantum Hall effect plateau transition,
where it provides a geometrical description of the disorder
in the Chalker-Coddington network model,
thus resolving the long-standing discrepancy
between numerically calculated and experimentally measured
localization-length exponents \cite{gkns, s-matrix}.

\section{Random surface ensembles}

Different surface randomization procedures generate statistically distinct ensembles,
each characterized by its own geometric and spectral properties.
The choice of randomization scheme therefore plays a central role in determining
the behavior of the resulting surfaces and the observables defined on them \cite{Kostov-1989}.
In this section, we introduce the formalism that will be used throughout this work and
define the specific randomization procedures considered.
These constructions provide the framework for comparing how
different notions of randomness influence the structure and
properties of the generated surface ensembles.

\subsection{Representations of quadrangular surfaces}

\begin{figure}
    \centering
    \includegraphics[width=\linewidth]{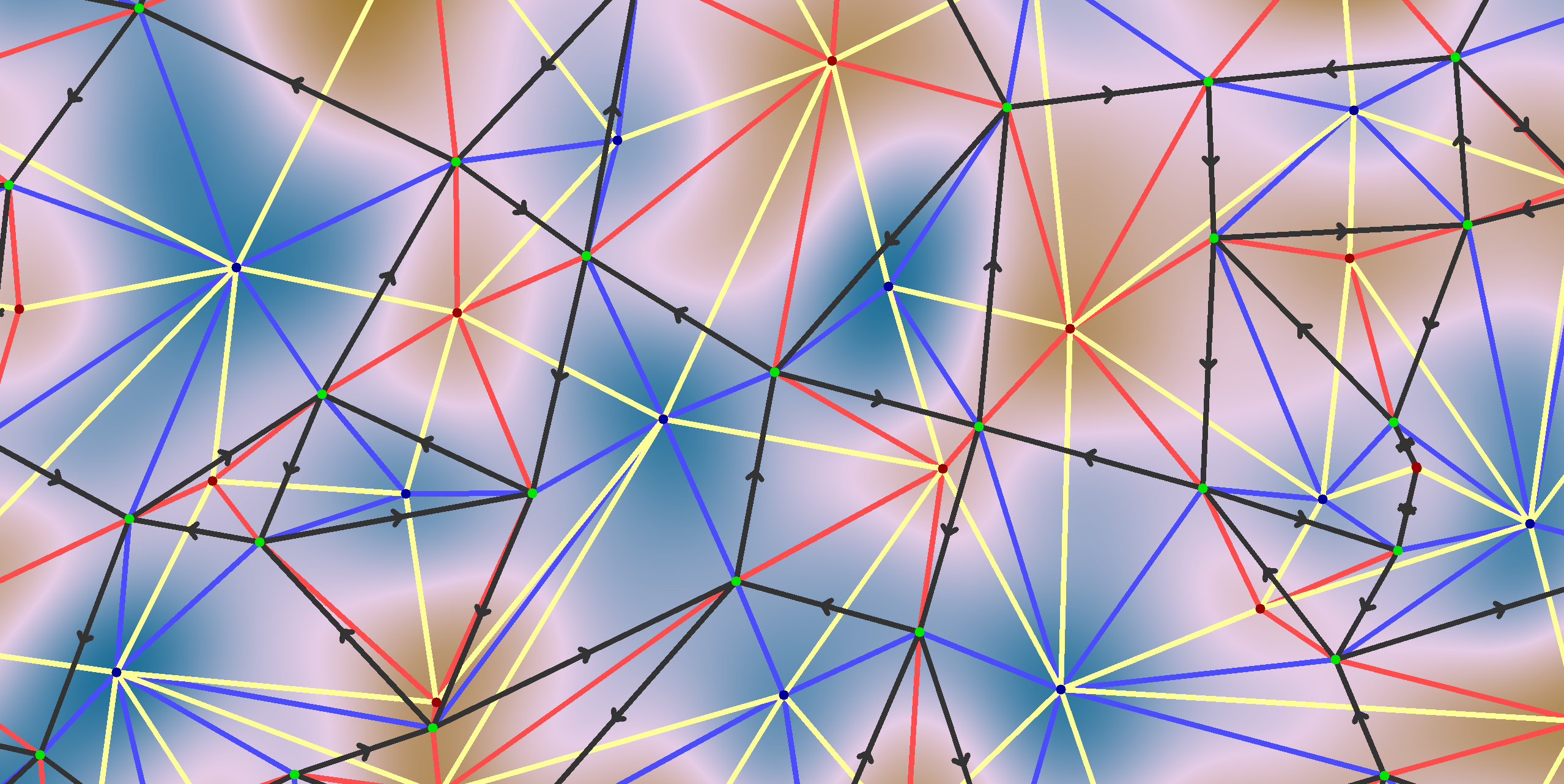}
    \begin{subfigure}[l]{.99\linewidth}
        \caption{
            Extrema surface (ES):
            all extrema as nodes,
            \Sp-\Min\ (blue) and \Sp-\Max\ (red) lines as links.
        }\caption{
            Scattering network (SN):
            \Sp-s as nodes, \Sp-\Sp\ (black) arrows as links,
            and \Min-s/\Max-s corresponding to faces.
        }\caption{
            The dual of SN (dSN):
            \Min-s and \Max-s as nodes,
            \Min-\Max\ (yellow) lines as links,
            and \Sp-s representing the faces.
        }\caption{
            Percolation lattice (PL):
            \Min-s as nodes, polylines \Min-\Sp-\Min\ (blue) as links,
            and \Max-s representing the faces, along with its dual, i.e.
            \Max-s as nodes, polylines \Max-\Sp-\Max\ (red) as links,
            and \Min-s as the faces.
        }
    \end{subfigure}
    \caption{The representations of random surfaces.
    A disordered potential
    (color map: increasing from blue to brown) with marked
    extrema points (\Sp: green, \Min: blue, \Max: red), and connections
    \Sp-\Min\ (blue lines), \Sp-\Max\ (red lines),
    \Min-\Max\ with a common \Sp\ (yellow lines) and
    \Sp-\Sp\ with a common \Min\ or \Max\ (black arrows).}
    \label{fig:rand_repr}
\end{figure}

We start with the discrete surface that naturally emerges in a
disordered real-valued potential $V(r)$ defined on a two-dimensional surface:
for each saddle point (\Sp) of the potential, there are two minima (\Min) and two maxima (\Max) 
that naturally correspond to it
(they are determined by a gradient ``descent" in the principal directions of $V(r)$ at the \Sp,
downward for the \Min-s and upwards for the \Max-s).
By connecting all pairs \Sp-\Min\ and \Sp-\Max, we end up with
a planar quadrangular surface of extrema points, the extrema surface (ES).
Also, in the planar projection the links of the \Sp-s are in a staggered
order (\Min-\Max-\Min-\Max) when moving around the \Sp.
As a result, we have a quadrangular planar surface with faces
consisting of \Sp-\Min-\Sp-\Max\ nodes (see \reff{rand_repr}).
The degrees (number of neighbor nodes) of the \Sp\ nodes are necessarily $4$,
and the \Min\ and \Max\ nodes can have arbitrary degrees.

From this picture, we can switch to the scattering network (SN) representation.
In this case, we switch each quadrangular face of ES for a directed link \Sp-\Sp.
The direction of the link is such that the \Min\ of the face is on its left.
We will end up with a network consisting of standard scattering nodes
(represented by the \Sp-s), and there will be a polygon in place of each \Min\ and \Max.
The degrees of all nodes are now $4$, and the number of sides of each polygon is equal
to the degree of the corresponding \Min\ or \Max.

A further node-to-face duality transformation will leave us with a
quadrangular ``dual scattering network" (dSN),
with its nodes representing the initial \Min-s and \Max-s of arbitrary degrees.
This will be the main representation used in this paper.

As a quadrangular surface, the dSN surface is inheritably bipartite,
with the nodes \Min\ and \Max\ as the two partitions.
We can now switch to the percolation lattice (PL) representation
by only linking the dSN representation nodes that had a common neighbor.
Hereby we end up with two disconnected planar surface partitions
defined by the \Min-s and \Max-s respectively,
that are node-to-face-dual to each other (\reff{rand_repr}).
The degree of each node is the same as in dSN.
This representation will be used in the analysis of
the percolation-related phenomena.

\subsection{Elementary randomization operations}

We will primarily work with the dSN representation, unless specified otherwise.
By the term ``distance" $d(p,q)$ between two nodes $p$ and $q$,
we will mean the minimum number of links
that one has to pass to reach one node from the other.
More formally, $d(p,p)=0$, $d(p,q)=1$ iff $p$ and $q$ are neighbors
and $d(p,q)=\min_r(d(p,r)+d(r,q))$.

In dSN, any two nodes at distance $2$ will be called a ``diagonal" $\eta$,
even if the nodes are not on the same quadrangle. In a quadrangle,
the notation $\bar{\eta}$ will be used for the diagonal opposite to $\eta$ (see \reff{gkns}).
The number of quadrangles will be referred to as the surface area $S$.

\begin{figure}
    \centering
    \begin{subfigure}[l]{.99\linewidth}
        \begin{tblr}{ccc}
            \SetCell[r=2]{}\includegraphics[width=.3\textwidth]{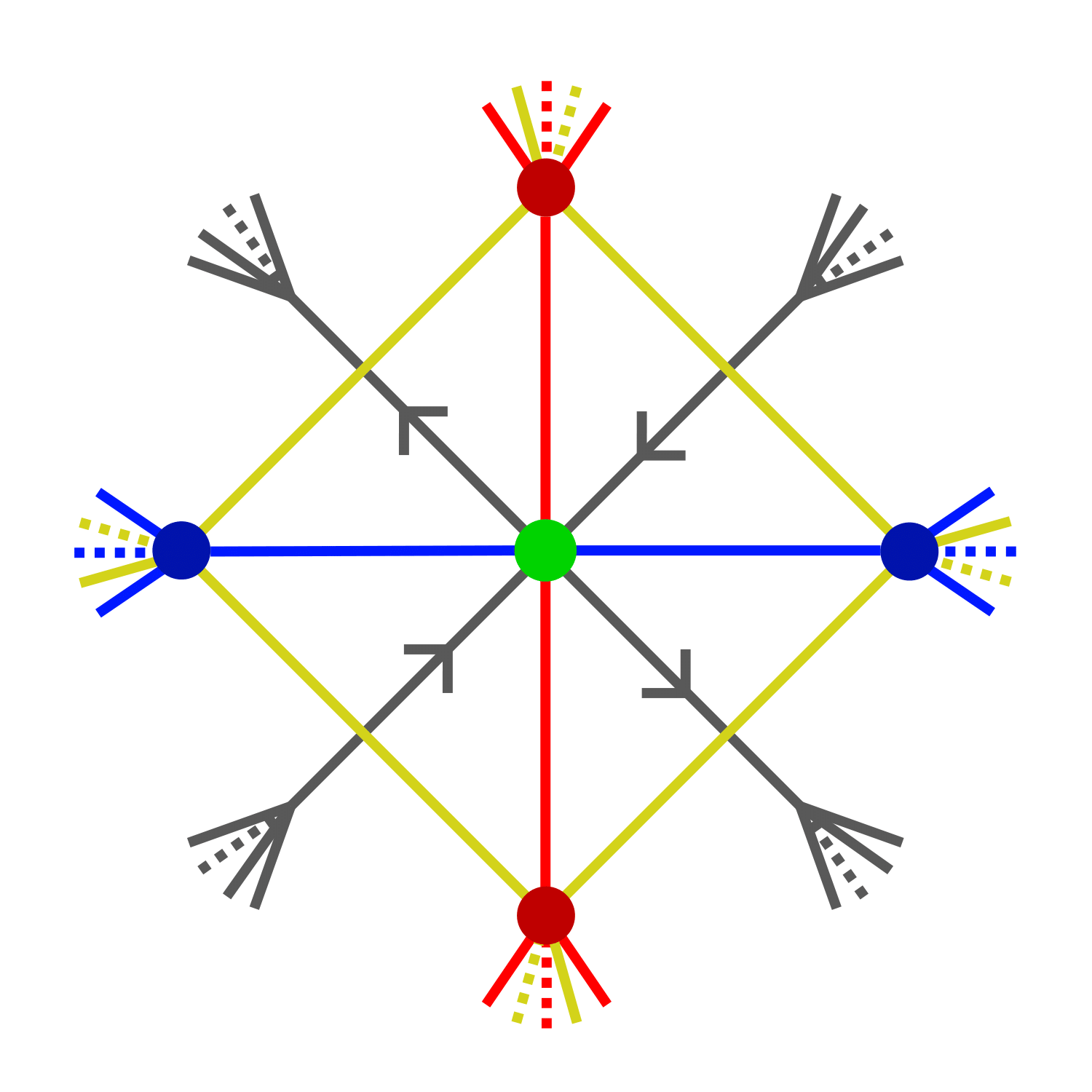}&
            \SetCell[r=2]{} $A_\eta$ \newline $\rightleftarrows$ \newline $A_\eta^\dagger~~$&
            \SetCell[r=2]{}\includegraphics[width=.3\textwidth]{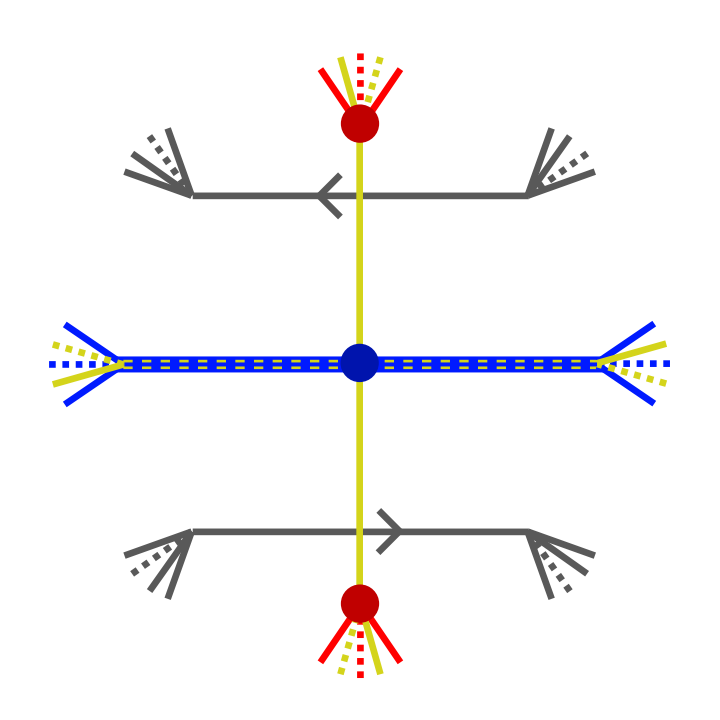}
            \\&&
        \end{tblr}
        \caption{The operation $A_\eta$ that generates the GKNS surface, and
            its inverse $A_\eta^\dagger$, with the two \Max\ points as $\eta$
            (and the two \Min\ points as $\bar{\eta}$).
            Joining two \Min-s by eliminating an \Sp\ (ES) or
            eliminating a scattering node (SN) or
            merging two nodes by diagonally squishing a quadrangle (dSN) or
            ``selecting" the \Min-\Min\ link (PL).
            A similar picture occurs if $\eta$ is the two \Min\ points.}
        \label{fig:gkns}
    \end{subfigure}
    \begin{subfigure}[l]{.99\linewidth}
        \begin{tblr}{ccc}
            \SetCell[r=2]{}\includegraphics[width=.3\textwidth]{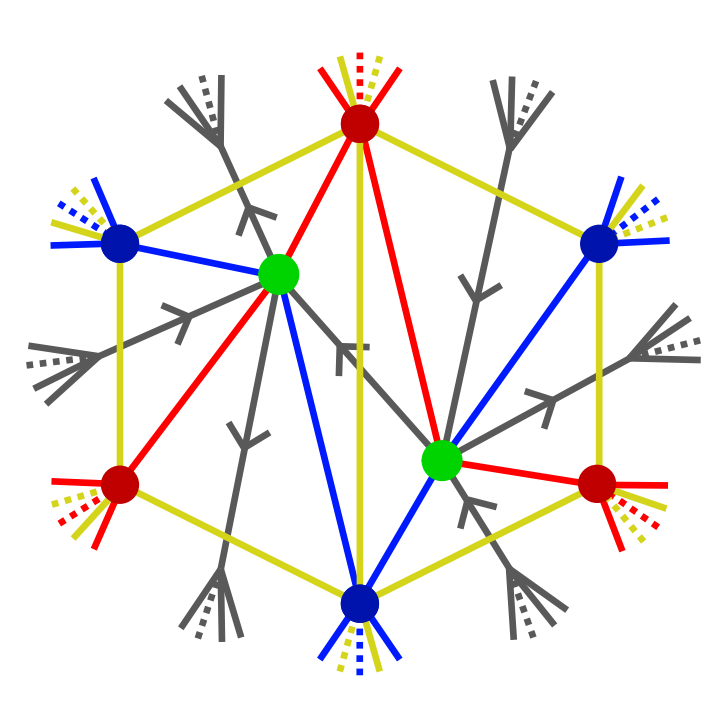}&
            \SetCell[r=2]{} $R$ \newline $\rightleftarrows$ \newline $R^\dagger~~$&
            \SetCell[r=2]{}\includegraphics[width=.3\textwidth]{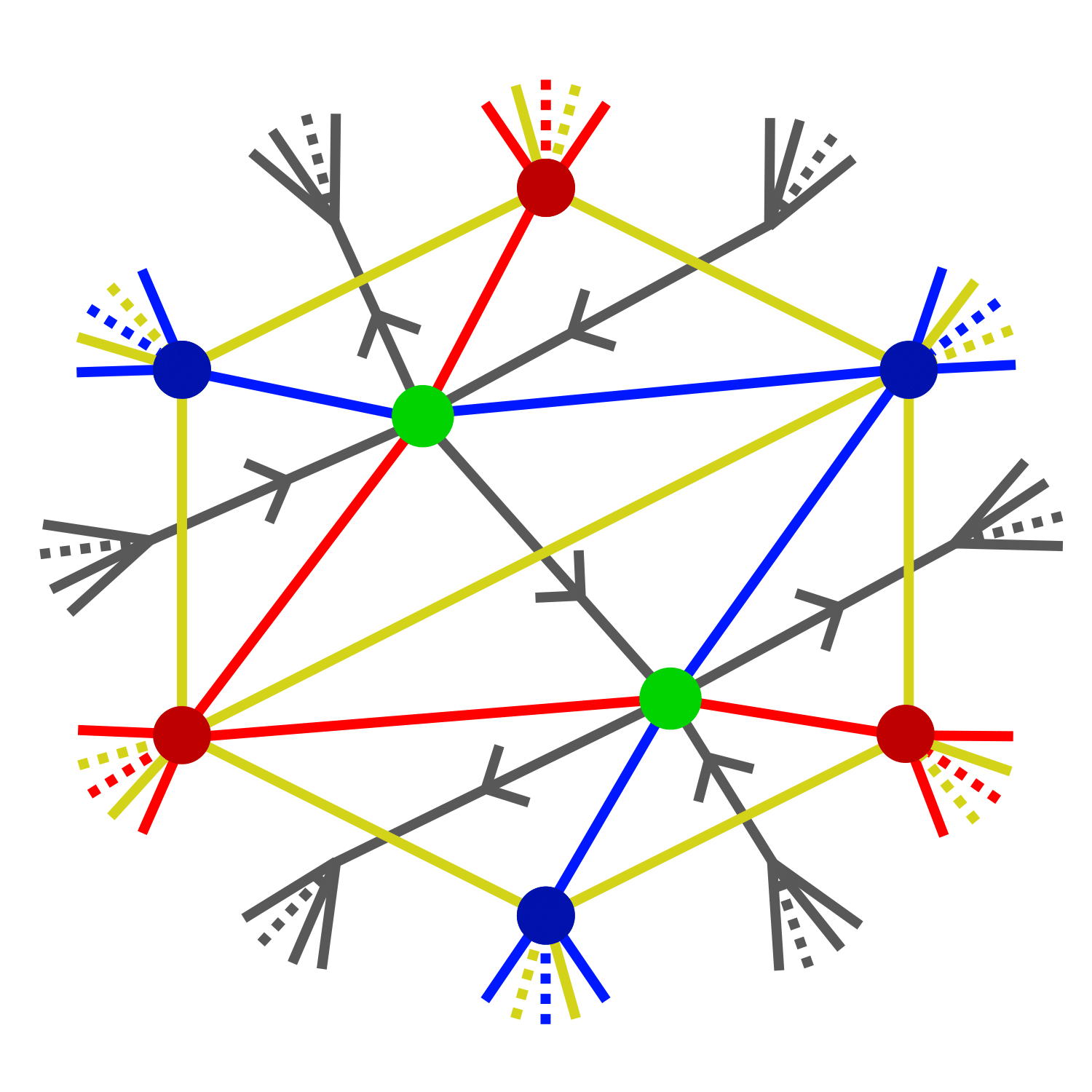} \\&&
        \end{tblr}
        \caption{The operations $R$ and $R^\dagger$ constituting
            the dynamical quadrangulation (DQ):
            rotation of a common link of two quadrangles (dSN)
            and the corresponding changes in the other representations.
        }
        \label{fig:dynam}
    \end{subfigure}
    \caption{The surface modifications in different representations.
        The notations are the same as in \reff{rand_repr}.
        The line segment fans 
        represent an arbitrary continuation of the surface.}
    \label{fig:modifs}
\end{figure}

\subsubsection{GKNS surface}
\label{sec:GKNS}

The GKNS random surface was defined \cite{gkns} in a study of the Integer quantum Hall effect
to properly capture the medium disorder \cite{gkns, conti}.
Replacing the regular network in the standard Chalker-Coddington model \cite{chal-codd}
by the GKNS network
solved the problem of discrepancy between the
theoretically (numerically) and experimentally obtained values
of the localization length exponent \cite{gkns, s-matrix}.
GKNS network is a modification of a regular scattering network by performing
a certain operation $A$ on each scattering node
equivalent to making the corresponding $2\times2$ scattering matrix
into an identity or an exchange matrix (see \reff{gkns}).
For each node, the operation is done with some probability $2p$,
in one of the two possible directions (equal probabilities for either direction).
Here we are only interested in connected networks,
and the operations that would split the network
into disconnected components are prohibited.
This adds a nonlocal constraint on the randomization procedure.
The definition of $p$ remains unchanged:
it is the probability that the operation $A$ is done at any given node,
in each direction.

In the dSN representation, the operation $A$ ``squishes" a quadrangle
by identifying two of its diagonally opposite points.
We index $A_\eta$ by the diagonal $\eta$ that remains not merged
after the operation (the identified nodes are those of $\bar{\eta}$).

In the PL representation, one can interpret the procedure of $A_\eta$'s action
as ``highlighting" the link $\eta$ and eliminating its dual $\bar{\eta}$ (\reff{gkns}).
The classical bond percolation is formulated as a lattice in which each possible link is
occupied (highlighted) with a given probability or is unoccupied otherwise \cite{percolation}.
The GKNS generation procedure can therefore be viewed as two competing percolations,
on the partitions \Min\ and \Max.
Additionally, the network connectedness restriction is transferred to this representation as
prohibiting a link highlighting that would result in a closed loop of highlighted links:
the scattering network would otherwise be split into
two disconnected components separated by the loop (see \reff{gkns}).
This constitutes the ``loopless" percolation \cite{arboreal-1}.
Thus, GKNS in the PL representation is
two mutually exclusive loopless percolation processes: 
the two processes are otherwise independent, except for the constraint that
whenever a link in one partition is occupied (highlighted),
its dual which belongs to the other partition can no longer be occupied (is eliminated).

The operation $A_\eta$ merges two nodes of $\bar{\eta}$ with degrees $n+1$ and $m+1$
into a single node of degree $n+m$,
and decreases the degrees of the two nodes of $\eta$ by $1$.
The surface area also decreases by $1$.
One can also define the operation $A_\eta^\dagger$ inverse to $A_\eta$,
which splits the $n+m$-degree node along the diagonal $\eta$
and thereby creates a quadrangle with diagonal $\eta$.

In this paper, we will refer to the surface with a specific value of $p$ as GKNS$_p$.

\subsubsection{Dynamical quadrangulation}

Here, we define dynamical quadrangulation (DQ) as an alternative way
of generating random quadrangular surfaces.
It is a generalization of dynamical triangulation (DT) \cite{Ambjorn-1985,David-1985,Kazakov-1985},
which is based on a link rotation operation on a triangular surface:
for any two adjacent triangles the shared link is removed
and replaced by the opposite diagonal.
The generalized link rotation operation $R$ for adjacent quadrangles
in dSN representation is shown in \reff{dynam}.
The surface area is invariant under this operation.
The random surface is produced by repeatedly randomly acting on the surface links
by the operations $R$ and its inverse $R^\dagger=R^2$, all with equal probabilities.

\subsubsection{Common basis}
\label{sec:com_bas}

The operations $A^\dagger$ can be classified by the degrees of the split nodes:
$A^\dagger$-s slitting an $n+m$-degree node into degrees $n+1$ and $m+1$
will be called $A_{(n)}^\dagger$ (or $A_{(m)}^\dagger$).

\begin{figure}
    \centering
    \begin{subfigure}[l]{.99\linewidth}
        \begin{tblr}{ccc}
            \SetCell[r=2]{}\includegraphics[width=.2\textwidth]{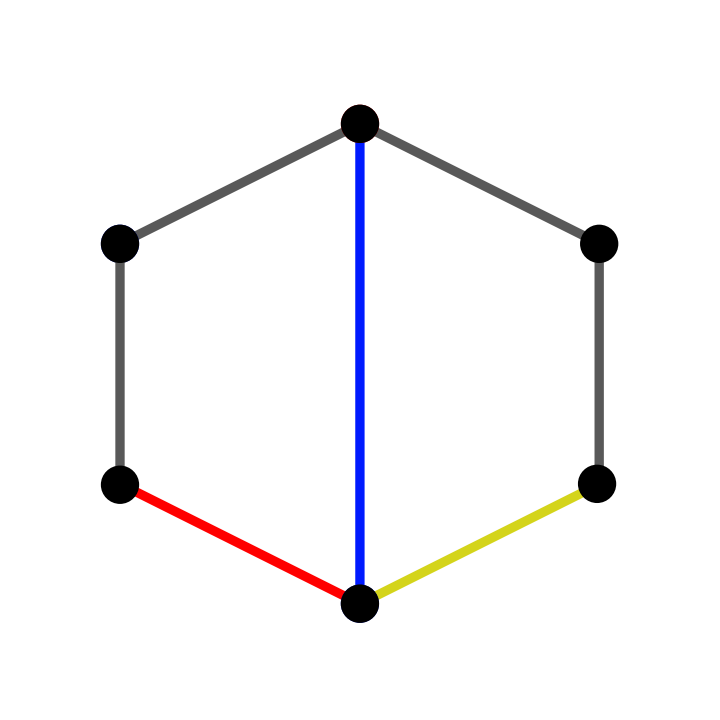}
            & \SetCell[r=2]{}$\xrightarrow[~^{~}_{~}]{A_{(2)}^\dagger}$ &
            \SetCell[r=2]{}\includegraphics[width=.2\textwidth]{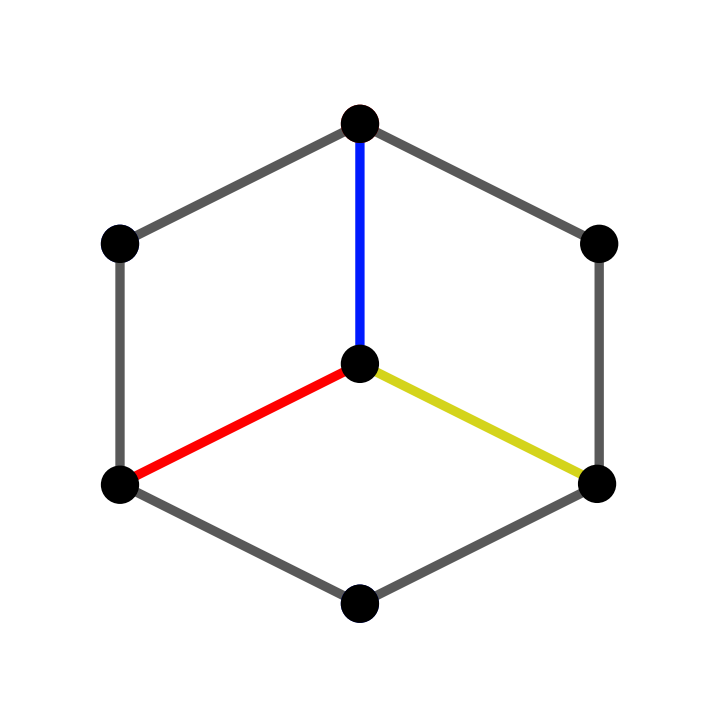}
            & \SetCell[r=2]{}$\xrightarrow{A}$ &
            \SetCell[r=2]{}\includegraphics[width=.2\textwidth]{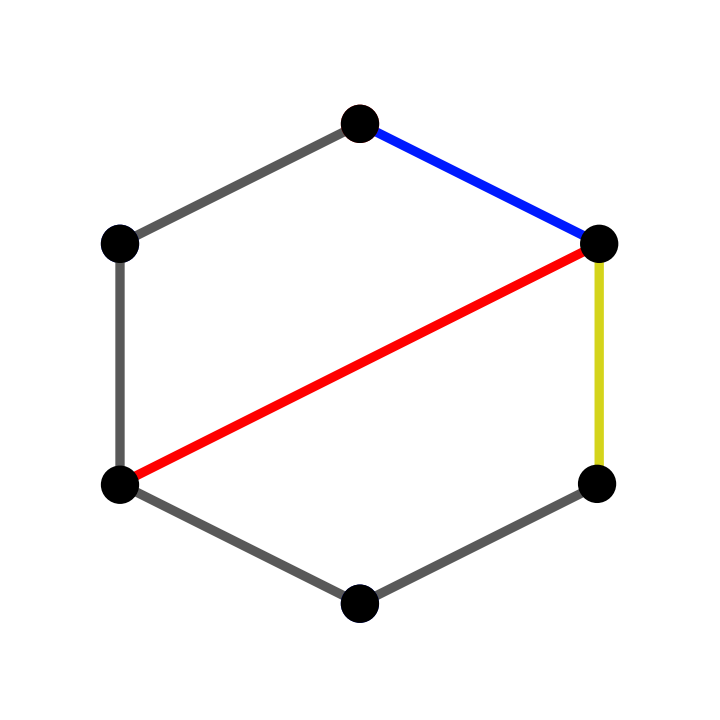} \\&&&&
        \end{tblr}
        \caption{The operation $R$ in terms of $A$ and $A_{(2)}^\dagger$.}
        \label{fig:raa}
    \end{subfigure}
    \begin{subfigure}[l]{.99\linewidth}
        \begin{tblr}{ccccc}
            \SetCell[r=2]{}\includegraphics[width=.2\textwidth]{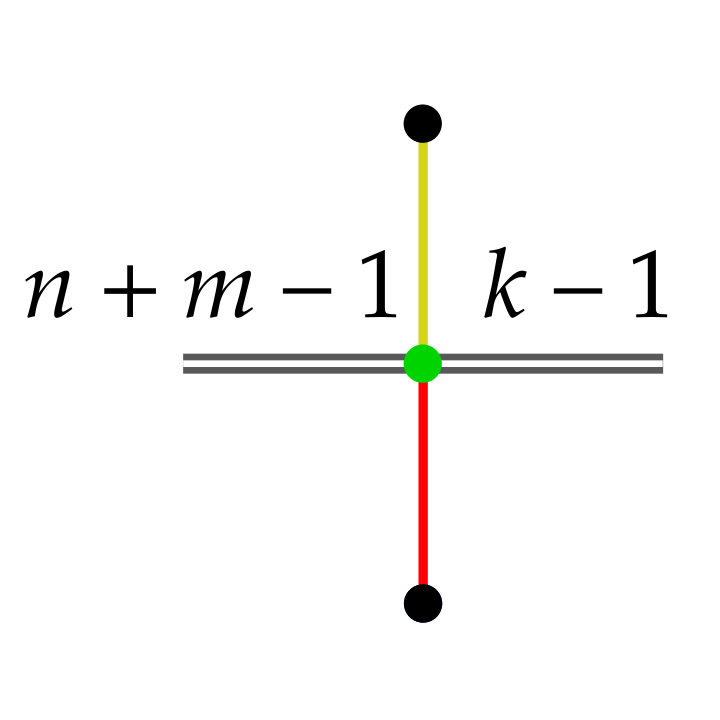}
            & \SetCell[r=2]{}$=$ &
            \SetCell[r=2]{}\includegraphics[width=.2\textwidth]{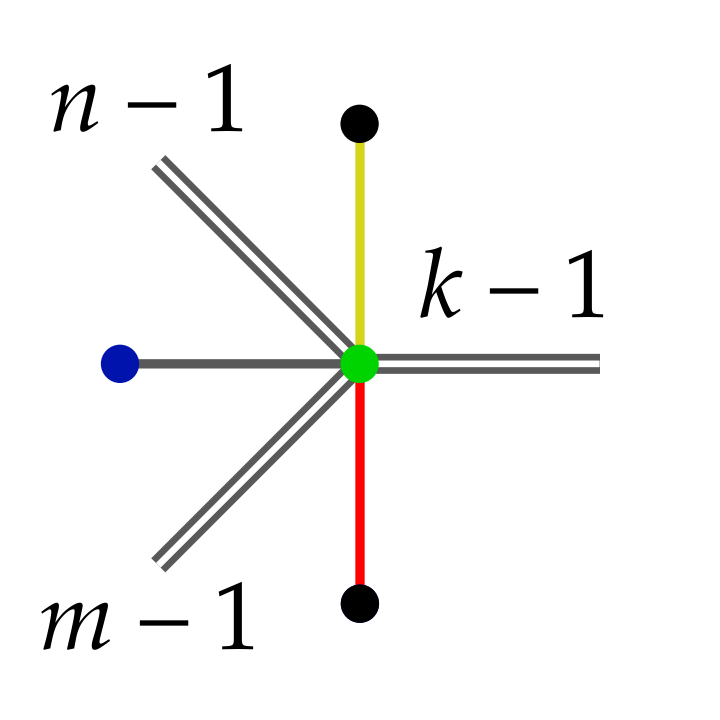}
            & \SetCell[r=2]{}$\xrightarrow[A_{(m)}^\dagger]{A_{(n)}^\dagger}$ &
            \SetCell[r=2]{}\includegraphics[width=.2\textwidth]{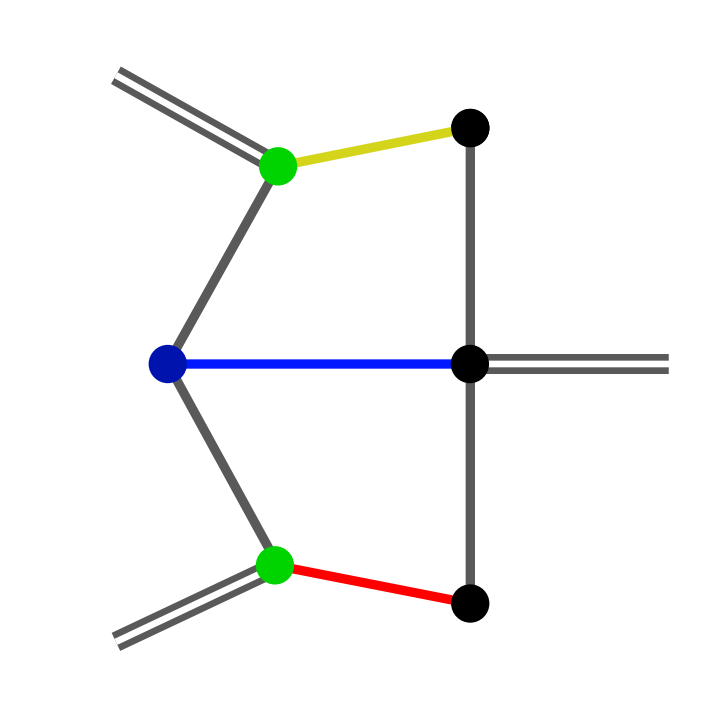} \\&&&&
            \\ \SetCell[c=2]{}$\downarrow A_{(n+m)}^\dagger \hspace{-16pt}$ 
            &&&& $\downarrow R \hspace{-4pt}$ \\
            \SetCell[r=2]{}\includegraphics[width=.2\textwidth]{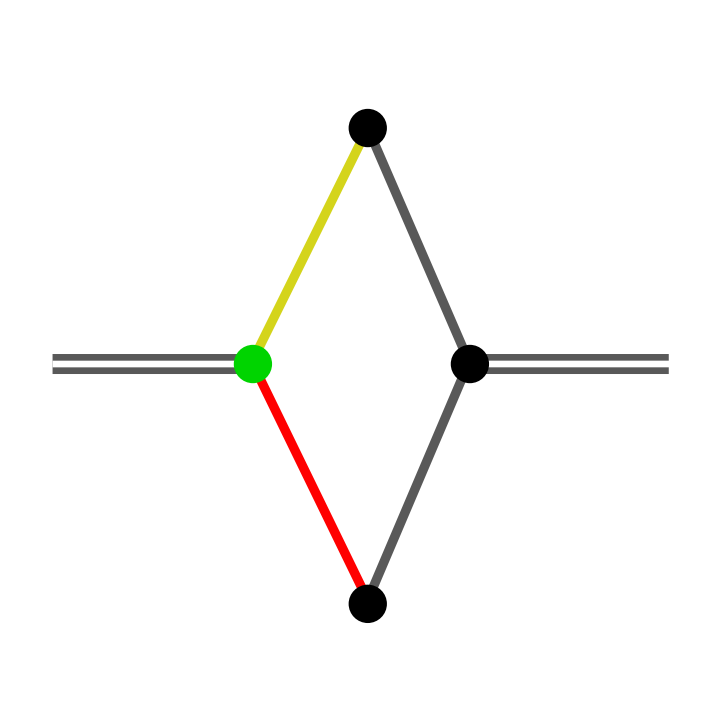}
            & \SetCell[r=2]{}$=$ &
            \SetCell[r=2]{}\includegraphics[width=.2\textwidth]{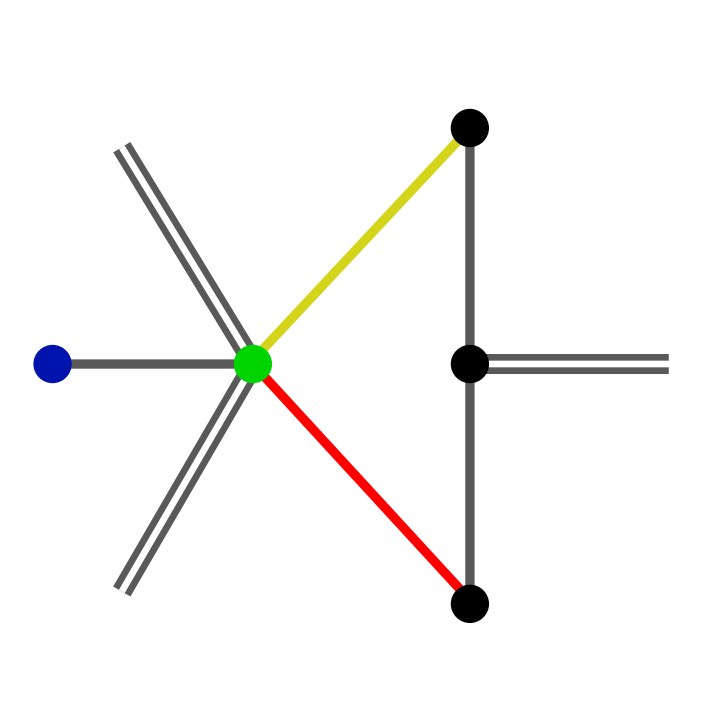}
            & \SetCell[r=2]{}$\xleftarrow{A}$ &
            \SetCell[r=2]{}\includegraphics[width=.2\textwidth]{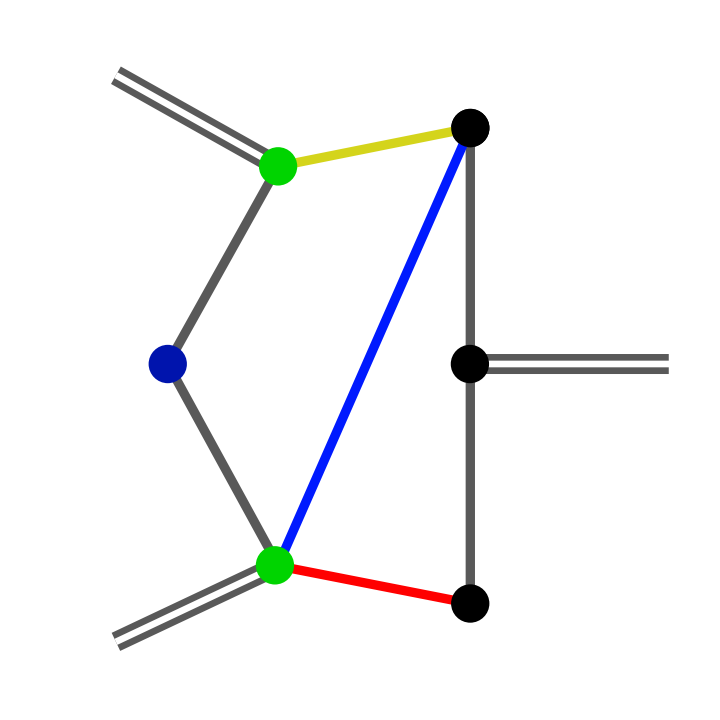} \\&&&&
        \end{tblr}
        \caption{The operation $A_{(n+m)}^\dagger$ in terms of
            $A_{(n)}^\dagger$, $A_{(m)}^\dagger$, $R$ and $A$.
            Double lines represent multiple links (the exact numbers are
            specified on the diagrams and are allowed be $0$).}
        \label{fig:adnm}
    \end{subfigure}
    \begin{subfigure}[l]{.99\linewidth}
        \begin{tblr}{cllcc}
            \SetCell[r=2]{}\includegraphics[width=.2\textwidth]{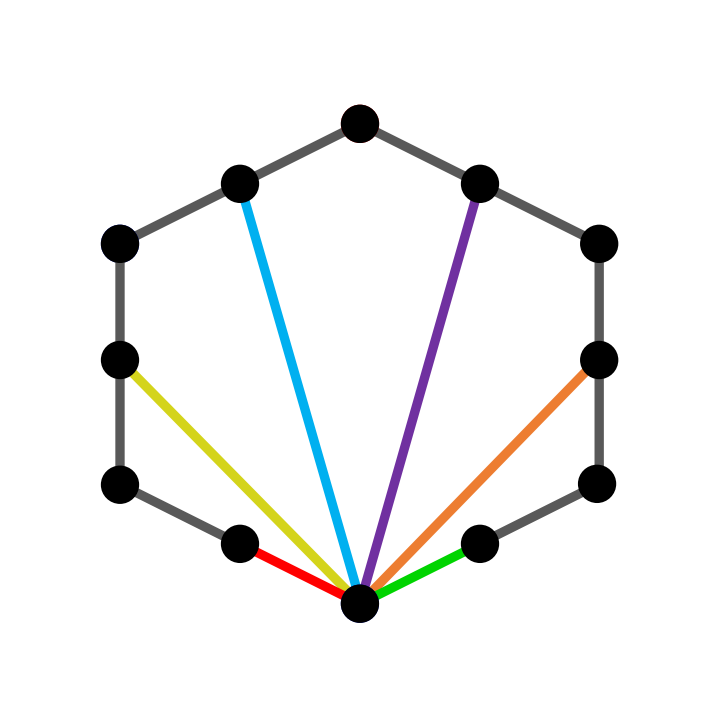}
            & \SetCell[r=2]{}$\xrightarrow[~^{~}_{~}]{A_{(k)}^\dagger}$ &
            \SetCell[r=2]{}\includegraphics[width=.2\textwidth]{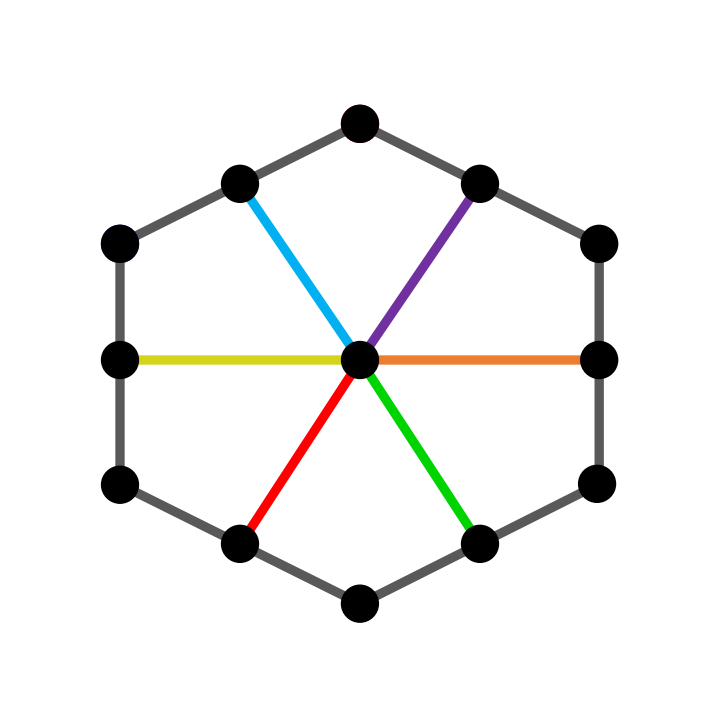}
            & \SetCell[r=2]{}$\xrightarrow[~]{A}$ &
            \SetCell[r=2]{}\includegraphics[width=.2\textwidth]{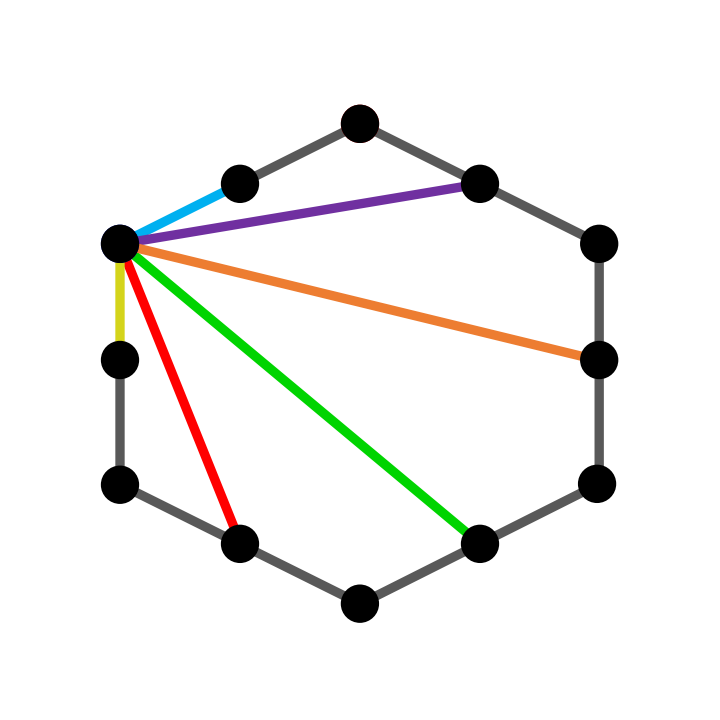} \\&&&&
            \\ $\downarrow R\hspace{-8pt}$ 
            &\SetCell[c=2]{}\downrightarrow $\cal R$&&& $||$ \\
            \SetCell[r=2]{}\includegraphics[width=.2\textwidth]{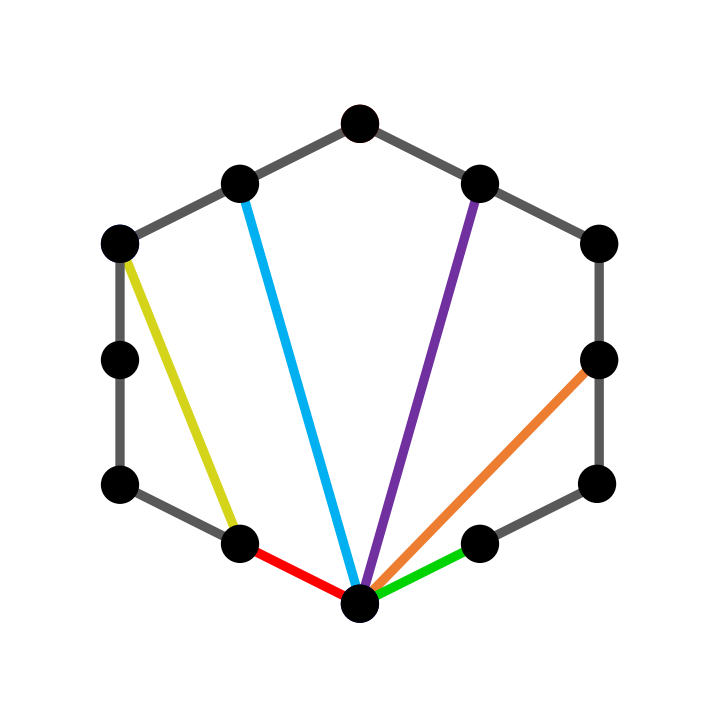}
            & \SetCell[r=2]{}$\xrightarrow[~]{R\cdots}$ &
            \SetCell[r=2]{}\includegraphics[width=.2\textwidth]{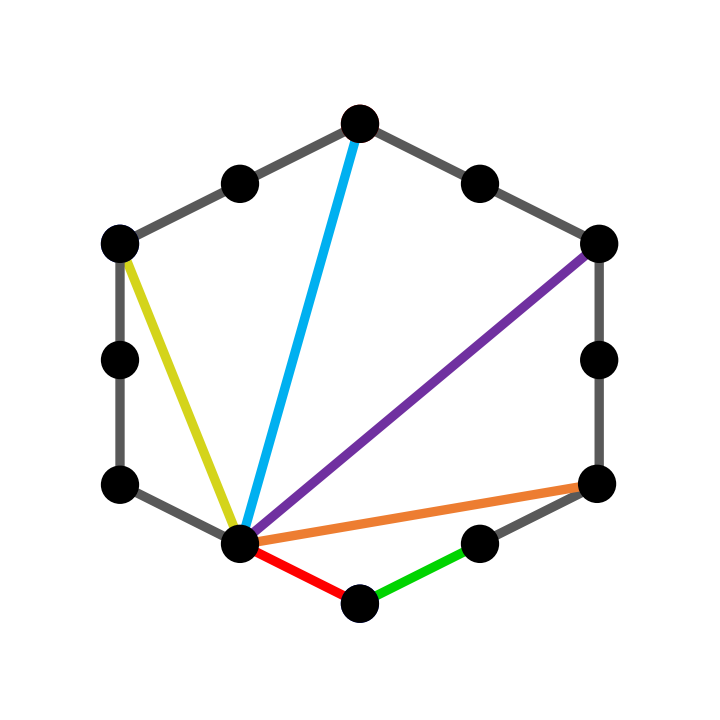}
            & \SetCell[r=2]{}$\xrightarrow[~]{\cal R\cdots}$ &
            \SetCell[r=2]{}\includegraphics[width=.2\textwidth]{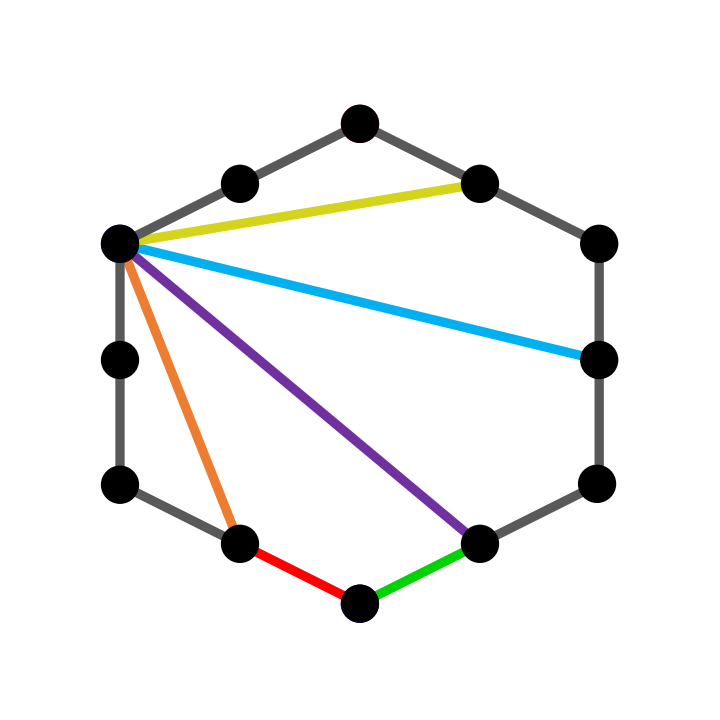} \\&&&&
        \end{tblr}
        \caption{The operations $A_{(k)}^\dagger$ and $A$ with a common node
            in terms of several series $\cal R$,
            which are $k-1$ consequent $R$-s. The example is for k=5.}
        \label{fig:aarr}
    \end{subfigure}
    \caption{The decompositions of operations $A^\dagger$, $R$ and $A A^\dagger$
        in dSN representation.
        Colors are there for the ease of following the procedure and don't represent anything.}
    \label{fig:basis}
\end{figure}

It can be shown that operations $A$, $A_{(n)}^\dagger$ and $R$ can all be expressed
just through $A$ and $A_{(2)}^\dagger$.
Obtaining $R$ as $A A_{(2)}^\dagger$ is shown in \reff{raa} and
the decomposition of $A_{(n+m)}^\dagger$ through
$A_{(n)}^\dagger$, $A_{(m)}^\dagger$, $R$ and $A$ is shown in \reff{adnm}.

Any $A_{(2k)}^\dagger$ can be split recursively. Formally,
\begin{equation}\nonumber
A_{(2k)}^\dagger \rightarrow A R A_{(2)}^\dagger A_{(2k-2)}^\dagger \rightarrow \cdots
\rightarrow A R A_{(2)}^\dagger \dots A R A_{(2)}^\dagger A_{(2)}^\dagger
\end{equation}
As $R$ itself is decomposed into $A$ and $A_{(2)}^\dagger$,
any $A_{(2k)}^\dagger$ can be expressed just through $A$ and $A_{(2)}^\dagger$.
Furthermore, for a node with an odd degree $2n+1$,
$A_{(2k+1)}^\dagger \equiv A_{(2n-2k)}^\dagger$,
thus covering $A_{(k)}^\dagger$-s for all values of $k$.
For an even degree node, the degree can be made odd (increased by $1$)
by the action of $A_{(2)}^\dagger$ (and later restored by $A$).
Thus, $A$ and $A_{(2)}^\dagger$ become generators for all operations on all possible nodes.

As we have seen (\reff{raa}), the consequent actions of $A_{(2)}^\dagger$ and $A$
such that the diagonals opposite to the ones on which they act have a common node
result in $R$ or $R^\dagger$.
Now we further show that any $A_{\eta'} A_\eta^\dagger$ such that
$\bar{\eta} \cap \bar{\eta}' \neq \varnothing$
can be expressed through $R$-s (\reff{aarr}).
If $A_\eta^\dagger$ is an $A_{(k)}^\dagger$ and the quadrangle created by it
is adjacent to the one to be collapsed,
$2(k-1)$ actions of $R$ (or $R^\dagger$) are required.
The number of actions becomes $4(k-1)$ if they have a common adjacent quadrangle,
$6(k-1)$ if there are two intermediate adjacent quadrangles, etc..

\subsubsection{General quadrangulation}

\begin{figure}
    \centering
    \includegraphics[width=.9\linewidth]{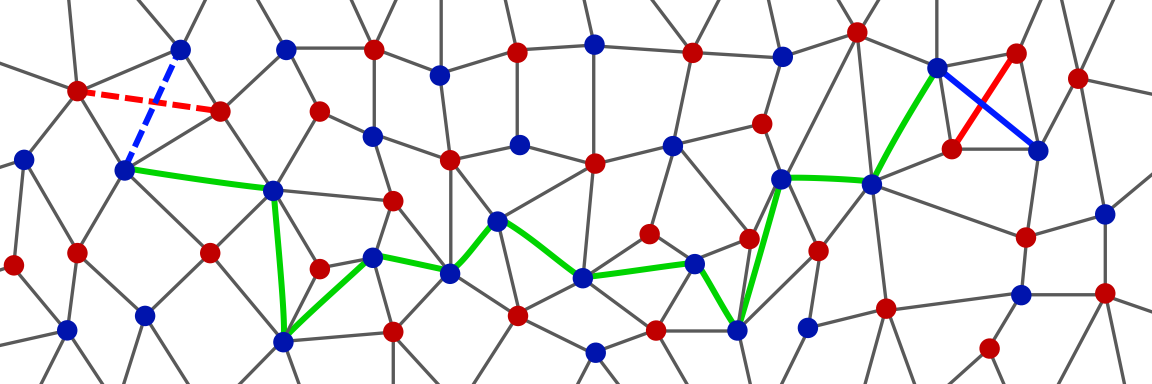}
    \caption{The bipartite (red and blue points) dSN surface (black lines),
        the diagonals $\eta_0$ (dashed) and $\eta'$ (solid) (red lines),
        the corresponding opposites $\bar{\eta}_0$ and $\bar{\eta}'$ (blue lines)
        and an example of a chain $\{\bar{\eta}_i\}$ continuously connecting them (green lines).}
    \label{fig:rchain}
\end{figure}

As we shall see later, the GKNS and DQ surfaces are substantially different.
In an attempt to define a surface that is obtained via area preserving operations as in DQ
but also preserves the properties of GKNS, we further define general quadrangulation (GQ).
As we have already seen (\reff{raa}),
the operation $R$ of DQ can be expressed as a combination of $A_{(2)}^\dagger$ and $A$,
and that $A_{(k)}^\dagger$ and $A$ are the complete classes of operators inverse to each other.
In GQ, we allow $A_{(k)}^\dagger$ and $A$ to act independently
at different places on the surface.
The quadrangle's diagonal $\eta$ for $A_\eta$ is selected randomly.
For $A_\eta^\dagger$, $\eta$ is selected by first randomly choosing a node,
then randomly choosing two of its neighbors to form $\eta$.
This pair of operations preserves the surface area.

In some cases, it is possible to describe the pair of operations $A_{\eta'} A_{\eta_0}^\dagger$
through a string of operations $R$.
Suppose that the diagonals $\eta_0$ and $\eta'$ belong to the same partition
of the bipartite dSN surface.
Then there are various connected diagonal chains $\bar{\eta}_0, \bar{\eta}_1, \dots, \bar{\eta}_L$, 
$\eta_L=\eta'$, $\bar{\eta}_i \cap \bar{\eta}_{i+1} \neq \varnothing$ (\reff{rchain}). Obviously,
\begin{equation}
    A_{\eta'} A_{\eta_0}^\dagger = A_{\eta_L}
    \inb({\prod_{i=1}^{L-1} A_{\eta_i}^\dagger A_{\eta_i}}) A_{\eta_0}^\dagger
     = \prod_{i=1}^{L} A_{\eta_i} A_{\eta_{i-1}}^\dagger \squad.
\end{equation}
The first equality is trivial, since $A_\eta^\dagger A_\eta = I$
(destroying a quadrangle and immediately creating it back).
As we know that the terms $A_{\eta_i} A_{\eta_{i-1}}^\dagger$ can be expressed through $R$-s,
then any $A_{\eta'} A_{\eta_0}^\dagger$ with $\eta_0$ and $\eta'$ on the same partition
can be expressed as a chain of $R$-s.

\subsection{Generation of surface ensembles}

As it was already mentioned, the GKNS surface ensemble is defined
starting with a regular surface,
and randomly applying the operation $A$ on the quadrangles
meanwhile conforming to connectedness constraints.
The probability of applying $A$ is $p$ in either of the two possible directions
and the quadrangle is left unmodified with probability $1-2p$.

Starting with the regular configuration as well,
DQ and GQ rely on a recurrent operation to obtain the surface ensemble,
unlike GKNS.
The approach is commonly known as Markov chains.
Obviously, some minimal number of operations is required to mimic the
ensemble at infinite number of operations.
A comprehensive analysis of the ensemble's convergence is therefore required.

In this subsection, we will discuss the DQ and GQ Markov chains and
the associated properties.

\subsubsection{Markov chains}

A Markov chain \cite{markov_chains} is the process of obtaining a random ensemble through
repeatedly acting on some state by a predefined set of operations.
It is a method of sampling from an ensemble of an unknown measure
over some space by defining its invariance properties.
With a given set of operations, their repeated action on
some initial state usually (see \refs{ergodicity}) leads to an equilibrium ensemble.
This ensemble is then invariant under the action of the operations and is independent of the initial state.

In our case, we considered the space $\mathcal G$ of
2-dimensional planar quadrangular surfaces of a given area $S$.
Then we define a set $O=\{(o, p_o)\}$ of operations
$o : \mathcal{G} \rightarrow \mathcal{G}$
with the corresponding probabilities $p_o, \sum_o p_o = 1$.
The ensemble $G$ is simply information about the probabilities $\{p_g\}$ of
getting the element $g\in\mathcal{G}$, $\sum_g p_g = 1$.
A step of a chain is the action of the set of operations on the ensemble
(or equivalently, the action of a randomly chosen operation on the space element),
\begin{equation}
    G^{(t+1)} = O G^{(t)} \squad\Leftrightarrow\squad g_{t+1} = o(g_t) \squad (o\in O )
\end{equation}
and then $p_g^{(t+1)} = \sum_{o(g')=g} p_o p_{g'}^{(t)}$.
The goal is to effectively reach the equilibrium (invariant) ensemble
$O G^{(\infty)} = G^{(\infty)}$.

In both the cases of DQ and GQ we start with the singular ensemble $G^{(0)}$
consisting only of one element $g_0$ (the regular surface of a given size),
$G^{(0)}(g_0)=1$.
For the DQ, the set of operations $O$ consists of the rotations $R^\pm$
of each link of the surface with equal probabilities.
For GQ, the operations are pairs of $(A, A_k^\dagger)$
acting independently at different positions on the surface.

\subsubsection{Ergodicity}
\label{sec:ergodicity}

The chains for which the equilibrium ensemble $G^{(\infty)}$ exists and is unique
are called ergodic. The chain is ergodic if it is irreducible
(all elements of the space can be connected by the chain)
and aperiodic (there is no $t_0 > 1$ such
that returning to the same element is possible only at multiples of $t_0$).

Generally, the discussed chains are not irreducible.
Obviously, the defined actions do not change the topology of the surface.
However, since we always start with the same $g_0$, the chain itself defines
the subspace (with the fixed topology of $g_0$) of accessible elements,
so it is irreducible per se.
The aperiodicity can be inferred from the discussion in Sec.\ref{sec:com_bas}.
In DQ, returning to the same element is possible both by 2 ($R, R^\dagger$)
and by 3 ($R$, $R$, $R$) operations on the same link.
The same is true for GQ: using the notation
$(A, A_\cdot^\dagger)_{ab}$ for the action of $A$ at $a$ and $A_\cdot^\dagger$ at $b$,
the 2-sequence ($(A, A_\cdot^\dagger)_{ab}$, $(A, A_\cdot^\dagger)_{ba}$) and the 3-sequence
($(A, A_\cdot^\dagger)_{ab}$, $(A, A_\cdot^\dagger)_{bc}$, $(A, A_\cdot^\dagger)_{ca}$)
both return to the same state if the positions $a, b, c$ are far enough so that
all operations are commutative.
Consequently, our chains are ergodic.

\subsubsection{Mixing and relaxation times}

Important properties of a Markov chain are its mixing and relaxation times
(the term ``time" refers to the number of chain steps).
The mixing time $t_{mix}$ of a chain characterizes the number of steps
required for the chain to be ``sufficiently close" to its equilibrium ensemble,
i.e. when $||G^{(t_{\text{mix}})}, G^{(\infty)}||$ becomes
small enough according to some notion of distance $||G^a, G^b||$ between the ensembles.
Usually $||G_1, G_2||$ is defined as the ``total variation distance"
which is $\text{TVD}(G^a, G^b)=\sum_g |p_g^a-p_g^b| / 2$ and
$\text{TVD} < 1/4$ is considered sufficiently small.
The mixing time is when the ensemble ``forgets" its initial state $g_0$.
The relaxation time is the rate of further convergence to equilibrium,
i.e. the rate of further decrease of $||G^{(t)}, G^{(\infty)}||$.
As we will see, it is the characteristic number of steps necessary to
acquire sufficient independence between successive samples.

However, this measure is not affordable, since the space of interest $\mathcal G$ is very large
and difficult to parametrize, and access to it is only through sampling.
For this reason, we will be studying these ensembles through ``observables".
We introduce the ``Network Laplacian Spectral Descriptors" (NetLSD) \cite{netlsd}.
These are based on the spectrum of the surface's normalized Laplacian matrix
defined as $\mathcal{L}_g=I - \mathcal{D}_g^{-1/2} \mathcal{A}_g \mathcal{D}_g^{-1/2}$
where $I$, $\mathcal{D}_g$ and $\mathcal{A}_g$ are the identity,
degree diagonal and adjacency matrices of the surface (graph) $g$, correspondingly.
In the dual (SN) representation $g^*$, all nodes have degree four,
so ${\mathcal D}_{g^*} \rightarrow 4I$ and $\mathcal{L}_{g^*}\rightarrow I-\mathcal{A}_{g^*}/4$.
NetLSD uses the Laplacian heat trace signature $h_\tau=\text{tr}\ e^{-\tau\mathcal L}$,
which in turn can be efficiently estimated by Stochastic Lanczos Quadrature \cite{slq}.
The smaller (bigger) values of $\tau$ reflect the local (global) structure of the surface.
The overall embedding system $g \rightarrow \mathcal L \rightarrow h_\tau$
provides a set of stable isomorphism invariant observables $h_\tau (g)$ for a given surface $g$.
The ``modes" $h_\tau$ can be separately analyzed for each $\tau$.

The mixing and relaxation times $t_\text{mix}$ and $t_\text{rel}$ can be studied
through observable deviations $\delta h_\tau=h_\tau-\inb<{h_\tau}>$, where
the notation $\inb<\bullet>$ stands for the average
over different chains and over times $t>t_0$.
The choice of thermalization (burn-in) time $t_0 \approx t_\text{mix}$
is such that the averaging is performed after the decay of the initial transients
and the chains are close to being stationary.
Defining $t_\text{mix}$ through $\delta h_\tau$ will introduce  a problem of self-consistency
($t_\text{mix}$ is used in the definition of $\delta h$ through $\inb<\bullet>$),
however it can be simply resolved iteratively.

\begin{figure}
    \centering
    \begin{subfigure}[l]{.49\linewidth}
        \includegraphics[width=\textwidth]{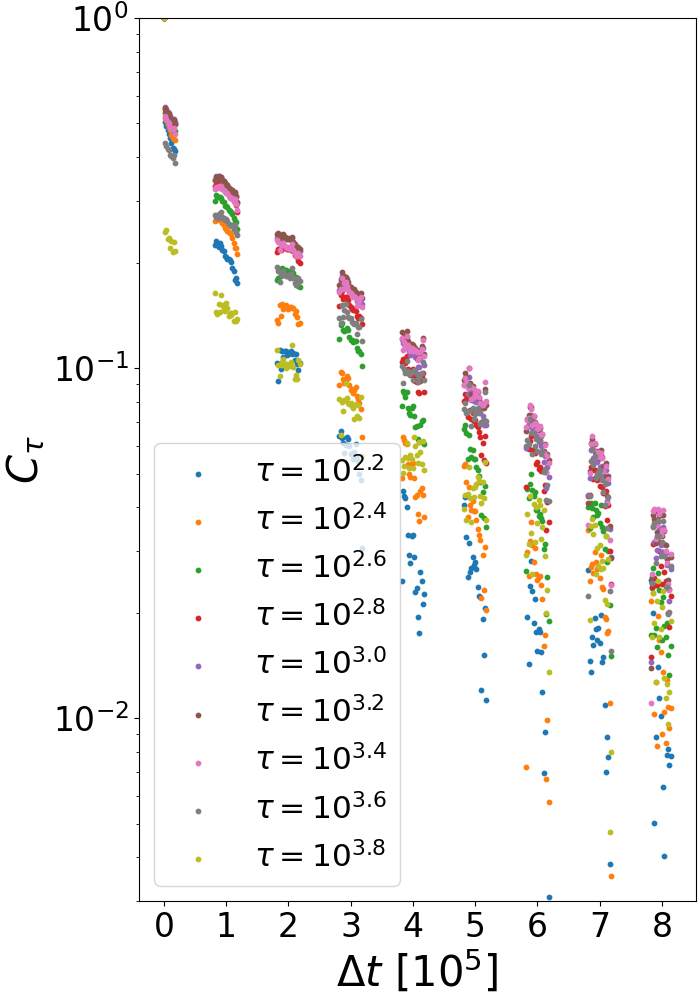}
        \caption{DQ}
    \end{subfigure}
    \begin{subfigure}[l]{.49\linewidth}
        \includegraphics[width=\textwidth]{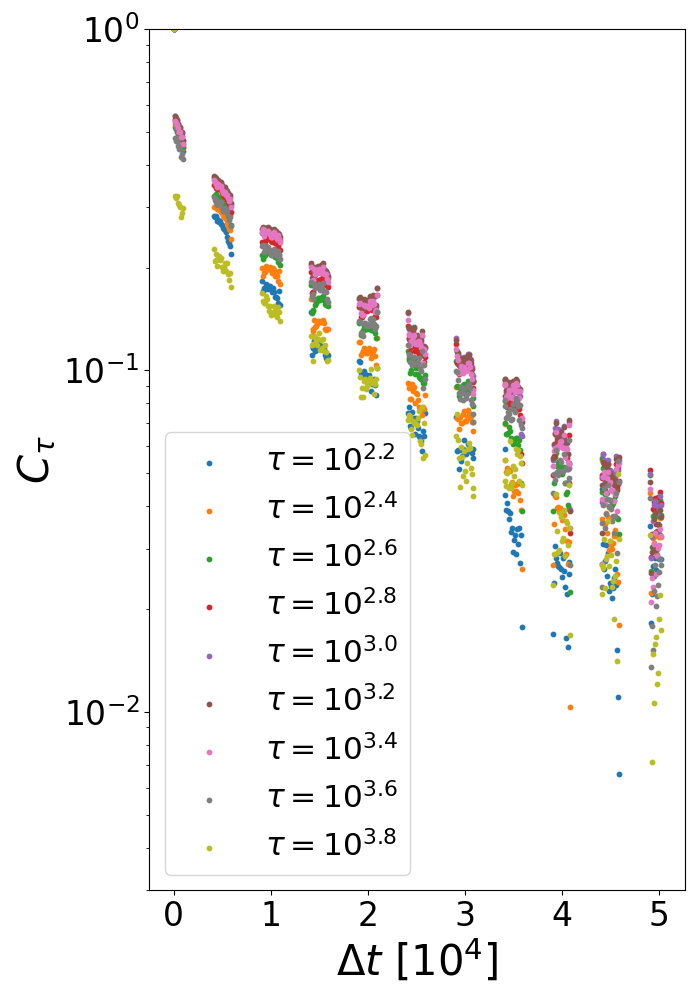}
        \caption{GQ}
    \end{subfigure}
    \caption{The dependence of the autocorrelation $C_\tau$ on the delay $\Delta t$
        for dynamical quadrangulation (DQ) and
        general quadrangulation (GQ) Markov chains,
        for various values of $\tau\in[10^{2.2}, 10^{3.8}]$
        in the case of surface area $S=100^2$.}
    \label{fig:autocorrs}
\end{figure}

The mixing time can be estimated as the number of steps required for
$\delta h_\tau(g_t)$ to reach the fluctuation scale $\sqrt{\inb<{\delta h_\tau^2}>}$
from its initial value of $\delta h_\tau(g_0)$.
To this end, for each chain $g(t)$ we measure the smallest time $t_c^{(\tau)}$
when $\delta h_\tau(g(t))$ crosses the value $0$,
i.e., becomes of opposite sign then $\delta h_\tau(g(0))$.
Then for each $\tau$ we take the $\eta=0.95$ quantile $Q_\eta$ over different chains $g$,
ensuring that at least the ratio $\eta$ of all realizations
are already within the fluctuation scale.
Afterwards, we maximize over parameters $\tau$,
as we are interested in the slowest mode $\tau$. Formally,
\begin{equation}
    t_\text{mix} = \max_{\tau} (Q_\eta[\min\{t\ |\ 
    \delta h_\tau(g(t))\cdot \delta h_\tau(g(0))<0\}]_g)
\end{equation}

For the relaxation time $t_\text{rel}$ we use
the autocorrelation functions \cite{autocorr} defined over chains $(g_0,g_1,\dots)$ as
\begin{equation}
    C_\tau(\Delta t) = \frac{\inb<{\delta h_\tau(g_{t+\Delta t}) \delta h_\tau(g_t)}>}
    {\inb<{\delta h_\tau^2(g_t)}>}
\end{equation}
This definition alone implies that the relaxation time
determines the minimum number of steps required between successive samples
to ensure sufficient independence.
The autocorrelation functions decay as
$C_\tau(\Delta t)\propto e^{-\Delta t/t_{\text{rel}}^{(\tau)}}$
where $t_{\text{rel}}^{(\tau)}$ is the relaxation time of mode $\tau$.
At large delays $\Delta t$ the autocorrelation is dominated by a single term
$C(\Delta t)\propto e^{-\Delta t/t_{\text{rel}}}$ and
the overall relaxation time $t_{\text{rel}}$ is defined by the slowest decaying mode $C_\tau$,
$t_{\text{rel}}=\max_\tau t_{\text{rel}}^{(\tau)}$.

\begin{figure}
    \centering
    \includegraphics[width=.7\linewidth]{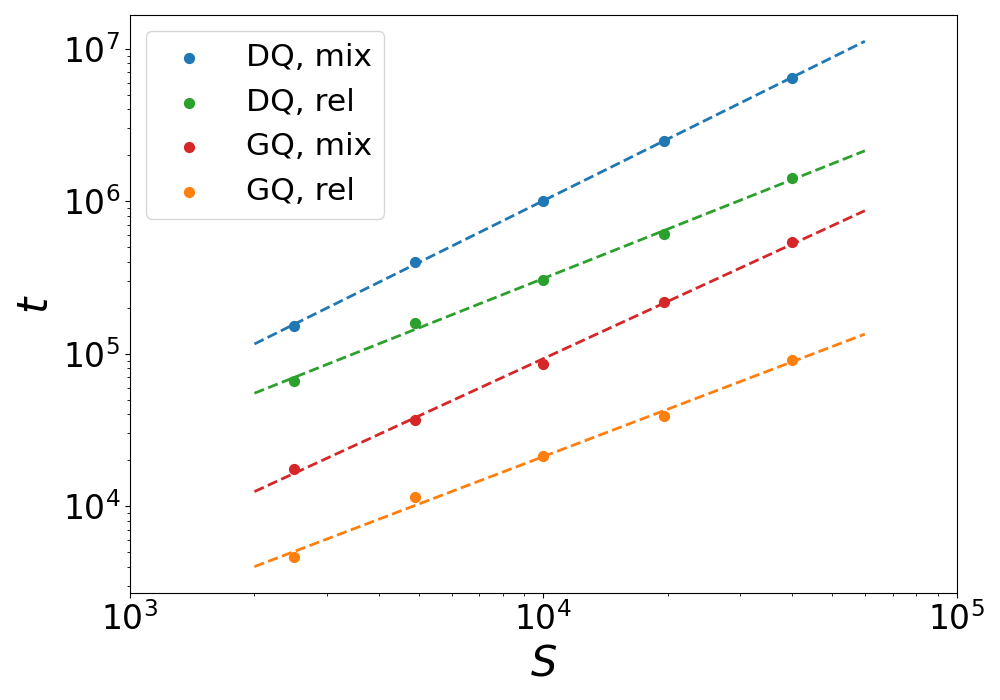}
    \caption{The scaling of mixing and relaxation times of DQ and GQ with surface area $S$.}
    \label{fig:t_scaling}
\end{figure}

According to both theoretical expectations and numerical results,
$h_\tau=0$ after a certain value of $\tau \gtrsim S$), so they do not carry any information.
The numerical calculations also show that the modes
that mix and relax the slowest for both DQ and GQ are those
with larger values of $\tau$ ($\sim S$), which describe the global structure of the surface.

The calculations have been performed for DQ and GQ separately,
on surfaces with areas $S=50^2, 70^2, 100^2,140^2, 200^2$,
over $250$ different chains each,
up to $t=4\cdot S^{3/2}$ steps per chain in the case of DQ and
$t=1 \cdot S^{3/2}$ steps per chain in the case of GQ.
The NetLSD data has been extracted for $51$ different values of $\tau\in[1,10^5]$
(the raw data is available in \cite{dataset}).
The typical behavior of $C_\tau(\Delta t)$ is shown in \reff{autocorrs}.
The extracted values of $t_\text{mix}$ and $t_\text{rel}$ for
both DQ and GQ depending on the surface area $S$
are shown in \reff{t_scaling}. The data implies scaling behaviors
\begin{equation}
\begin{split}
    t_\text{mix}^\text{(DQ)} \approx 4.2\cdot S^{1.34} \quad&,\quad
    t_\text{rel}^\text{(DQ)} \approx 15\cdot S^{1.08} \squad , \\
    t_\text{mix}^\text{(GQ)} \approx 0.95\cdot S^{1.25} \quad&,\quad
    t_\text{rel}^\text{(GQ)} \approx 1.6\cdot S^{1.03}
\end{split}
\end{equation}
with surface area $S$.

For proper calculations, 
the sampling of the chains has to be done at times $t \gtrsim t_\text{mix}$,
so that they are from the equilibrium ensemble,
and with time intervals $\Delta t \gtrsim t_\text{rel}$,
so that the samples can be considered independent.
We will be using $t \geq 2 t_\text{mix}$ and $\Delta t = 2 t_\text{rel}$.

\section{Characteristic properties}

In this section, we will explore the key properties (e.g. critical exponents)
of the defined random surfaces.
We will establish the distinctive features that differentiate them both from each other
and from other established models \cite{barabasi}.
All the generated data can be found in \cite{dataset}.

\subsection{Percolation properties of GKNS}

The percolation properties of quadrangular surfaces deal with how connectivity and
cluster formation emerge in the PL representation.
As already established in \refs{GKNS},
the generation of the GKNS$_p$ network is two mutually exclusive loopless percolation procedures
defined on regular lattices dual to each other,
where two intersecting links cannot be simultaneously occupied
and the occupation of either link has the same probability $p$ (\reff{percolation}).

\begin{figure}
    \centering
    \includegraphics[width=.9\linewidth]{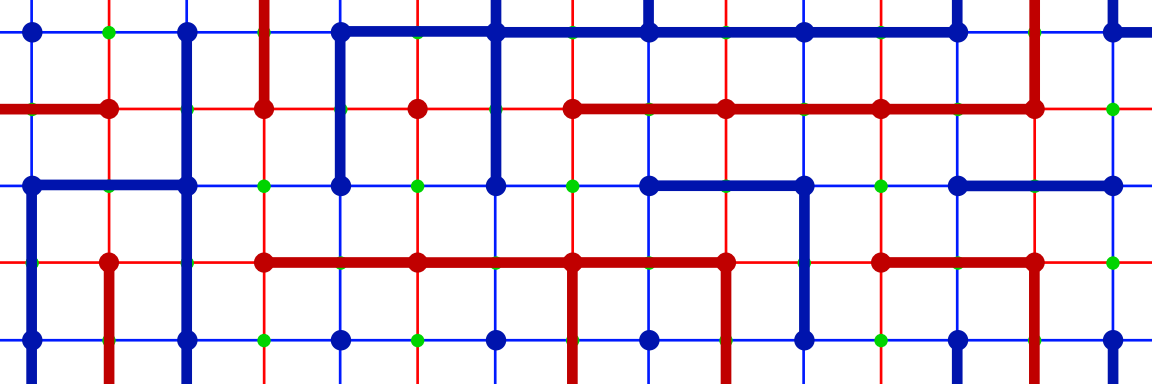}
    \caption{A GKNS realization on a regular network in the PL representation.
        The \Min\ (blue nodes and lines) and \Max\ (red nodes and lines) partitions,
        with the occupied (highlighted) links being depicted thickened.
        Green dots portray the \Sp-s of the initial ES representation.}
    \label{fig:percolation}
\end{figure}

It is important to note that in this mutually constrained construction of GKNS$_p$,
each of the partitions separately is no longer a loopless percolation with probability $p$.
This is possible, since in the overall ensemble the two partitions are locally correlated,
which, given the nonlocal constraints that exist within either partition,
induce a change in statistical properties.
The statement can be verified numerically by studying cluster formation
on surfaces of area as small as $S=36$.
This correlation makes the percolation picture of GKNS
substantially different from the loopless or regular percolation
(the two are known to have the same critical behavior \cite{arboreal-1, arboreal-2})
on a single lattice.

A cluster is defined as a set of nodes connected via occupied links.
We can assign coordinates $(x,y)$ to each node, based on their position in the regular lattice,
so that the coordinate difference of nearest neighbor nodes is $1$.
Then each cluster is enclosed by a bounding box whose boundaries are defined
by the minimum and maximum coordinates of its contained nodes.
In regular percolation on an infinite lattice,
an infinite cluster (a cluster with a divergent bounding box)
emerges at the critical point $p_{0,r}=1/2$ \cite{percolation, barabasi}.

Here we study the correlation length and its divergence near the critical point $p_0$,
which can be different from $1/2$ in our case.
Correlation length is defined through the function $F_d(d)$ which is
the probability that a given node is connected to (is in the same cluster as)
some node on a square of size $2d$ centered around it, or,
in other words, has a ``reach" of at least $d$.
$F_d(d)$ is expected to decay exponentially as $F_d(d)\propto e^{-d/\xi_p}$
where $\xi_p$ defines the correlation length \cite{barabasi}.
For each node, one can define its ``reach" $d$ as the distance
between the node and the farthest border of its cluster's bounding box.
The reach distribution function $f_d(d)$ is connected to $F_d(d)$ by
$f_d(d)\propto F_d(d)-F_d(d+1)$ (proportionality is to ensure normalization)
and thus decays similarly, as $f_d(d)\propto e^{-d/\xi_p}$.
The correlation length divergence is expected to have the form
$\xi_p \propto (p_0-p)^{-\nu}$ near the critical point $p\rightarrow p_0$.

Another quantity of similar nature is the characteristic cluster size $s_p$.
The size of a cluster $s$ is the number of points within the cluster.
Cluster size distribution $f_s(s)$ is defined as
the probability of a given point being in a cluster of size $s$. Similarly,
the distribution is expected to decay exponentially as $f_s(s)\propto e^{-s/s_p}$ for large $s$,
and $s_p$ is expected to diverge at the critical point
as $s_p \propto (p_0-p)^{-1/\sigma}$ \cite{barabasi}.
For numerical stability, we also introduce an analogous ``integral" form similar to $F_d$,
$F_s(s)\propto\sum_{s'\geq s} f_s(s')$,
which also decays in an exponent of rate $s_p$.

\begin{figure}
    \centering
    \begin{subfigure}[l]{\linewidth}
        \includegraphics[width=.49\linewidth]{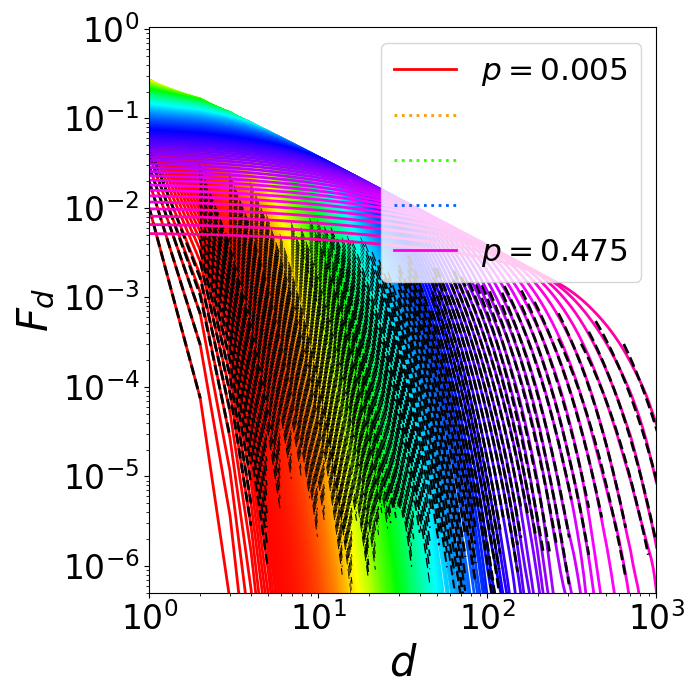}
        \includegraphics[width=.49\linewidth]{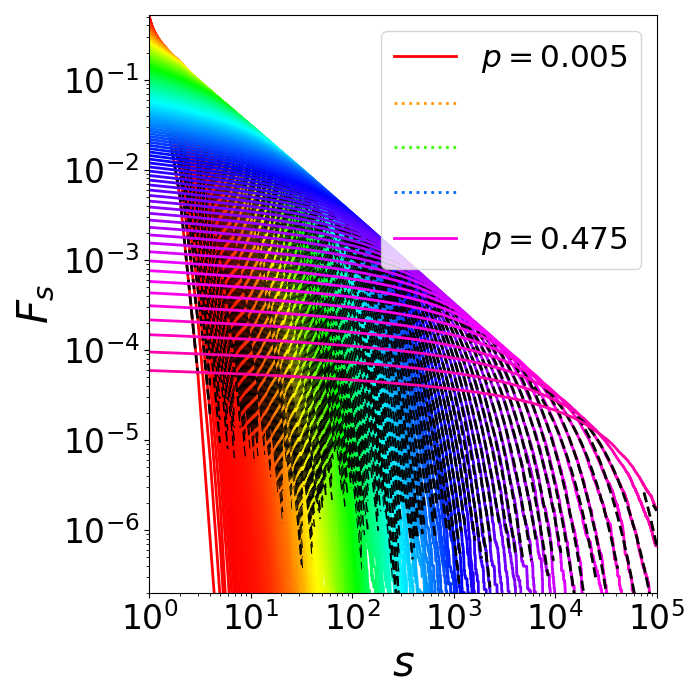}
        \caption{``Integral" reach distribution $F_d(d)$ (left) and
            cluster size distribution $F_s(s)$ (right) 
            (solid lines) for various values of $p$ (in different colors) with
            the corresponding fitting lines
            $F_d(d)\propto e^{-d/\xi_p}$ and $F_s(s)\propto e^{-s/s_p}$ 
            respectively (dashed lines), allowing to determine the values of
            $\xi_p$ and $s_p$ for a given $p$.}
        \label{fig:perc_raw}
    \end{subfigure}
    \begin{subfigure}[l]{\linewidth}
        \includegraphics[width=.49\linewidth]{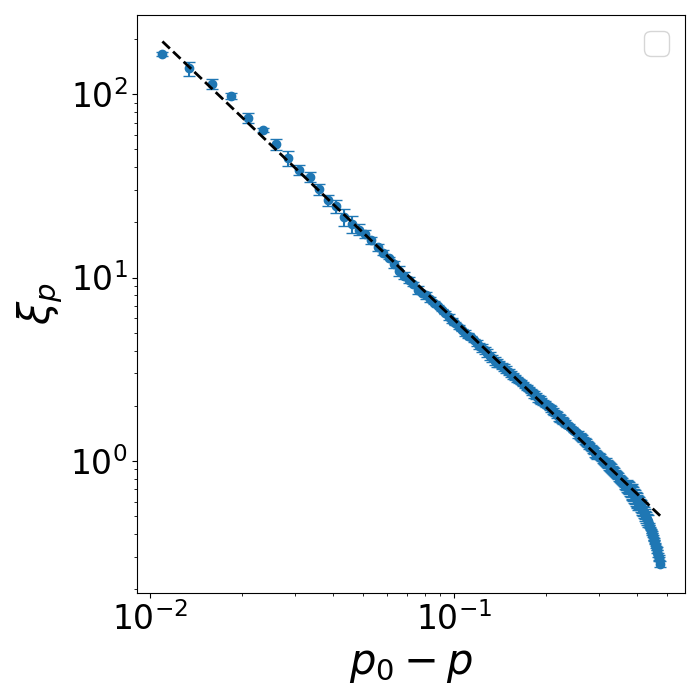}
        \includegraphics[width=.49\linewidth]{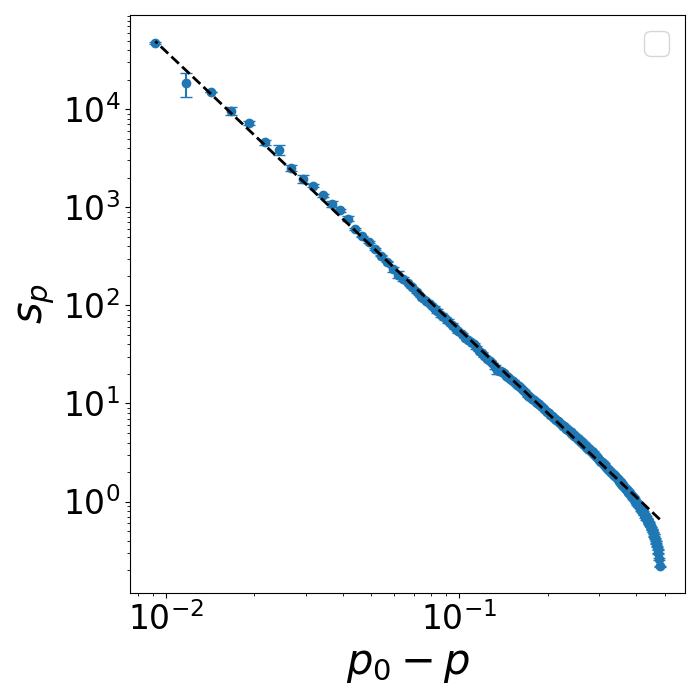}
        \caption{The dependence of $\xi_p$ (left) and $s_p$ (right) on $(p_0-p)$
            with the corresponding fitting lines
            $\xi_p\propto(p_0-p)^{-\nu}$ and $s_p\propto(p_0-p)^{-1/\sigma}$ (dashed lines),
            providing the critical point position $p_0=0.4851(13)$ and
            the exponent values $\nu=1.588(17)$ and $1/\sigma=2.83(2)$.}
        \label{fig:perc_ripe}
    \end{subfigure}
    \caption{The determination of critical point position $p_0$,
        correlation length exponent $\nu$ (left)
        and cluster size divergence exponent $\sigma$ (right),
        for the GKNS network in PN representation.
        All plots are in log-log scale.}
    \label{fig:perc_data}
\end{figure}

Numerical calculations for $\xi_p$ and $s_p$ are
performed on $100$ different open surface realizations
of area $S=L\times L$ with $L=5000$.
The data for $f_d$ and $f_s$ is extracted only from the middle
section of the surface with linear size $L'=2000$.
In each realization, values of $p$ with a step of $\Delta p=0.005$ are considered,
in the range for which none of the clusters that contain a point
from the sampled middle section touches (feels) the surface boundary.
The maximum considered value of $p$ common to all realizations is $0.475$.
Then it can be stated, that the data for lower values of $p$
are completely free of finite-size effects.
$\xi_p$ and $s_p$ are extracted by a fit of the averaged
$f_d$ and $f_s$ functions (\reff{perc_raw}),
which then can be used to determine $p_0$, $\nu$ and $\sigma$.
The later are calculated by a simultaneous two-curve fit
with a shared parameter $p_0$ (\reff{perc_ripe}).
The uncertainties are estimated using the covariance matrix
obtained from the least-squares fitting procedure.

The results show that the criticality is affected
by the correlation between the two (\Min\ and \Max) partitions,
however not in an obvious way.
The determined critical point position is $p_0=0.4851(13)$,
and the corresponding exponents are $\nu=1.588(17)$ and $1/\sigma=2.83(2)$ (\reff{perc_data}).
These values are stable against the change of the fitting range of $p$
in the region $p>0.08$.

The values of the mentioned exponents in regular percolation are
$\nu_r=4/3\approx 1.33$ and $1/\sigma_r=91/36\approx2.53$ \cite{barabasi,percolation}.
This implies that the correlation between the partitions \Min\ and \Max\ 
not only shifts the critical point position from $p_{0,r}=1/2$ to $p_0=0.4851(13)$,
but also modifies the criticality,
which is reflected in the slight change of critical exponents.
This implies that the GKNS percolation is in
a different universality class than regular/standard percolation.

\subsection{Geometric properties of GKNS, DQ and GQ}

In this section, we discuss the structural properties of the studied surfaces.
These reflect on the behavior of a model defined on a random surface \cite{Janke-2004, harris},
as well as define the surface geometry dynamics \cite{Polyakov-1981,Polyakov-book}.

\begin{figure*}[t]
    \centering
    \begin{subfigure}[l]{.245\linewidth}
        \includegraphics[width=\textwidth]{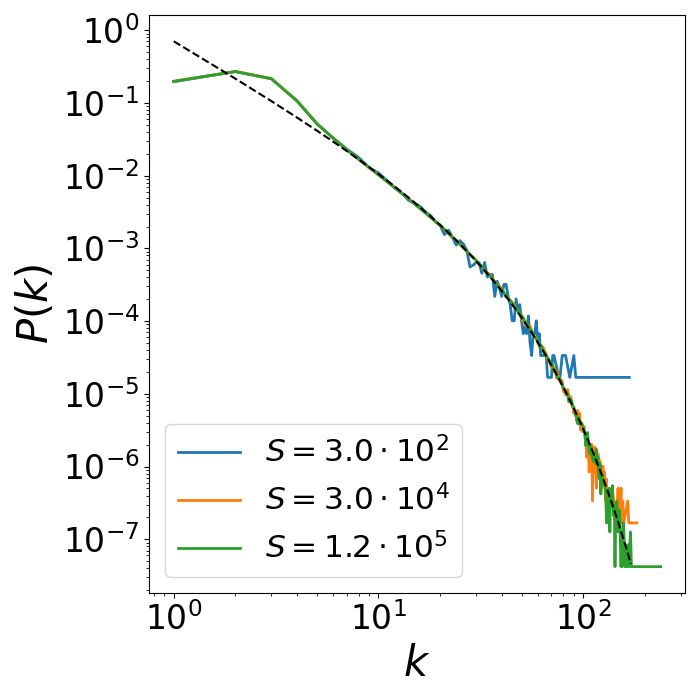}
        \includegraphics[width=\textwidth]{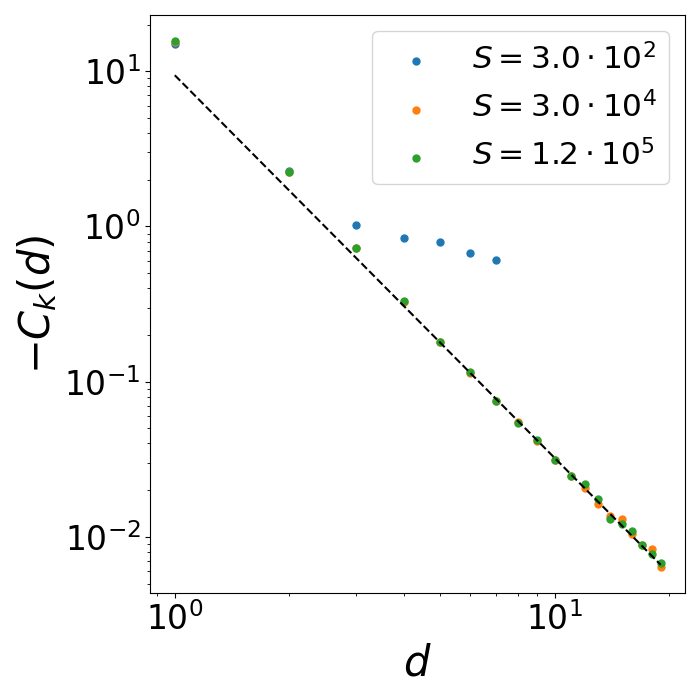}
        \includegraphics[width=\textwidth]{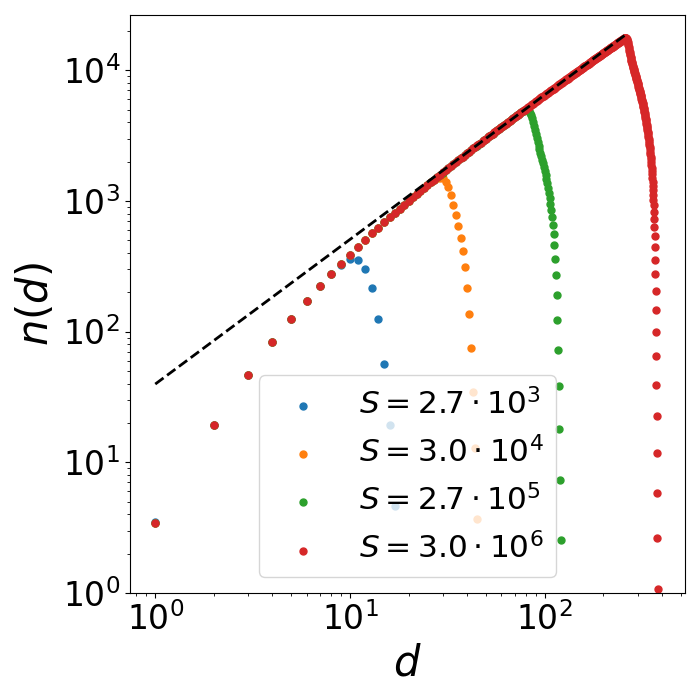}
        \caption{GKNS$_{1/3}$.}
    \end{subfigure}
    \begin{subfigure}[l]{.245\linewidth}
        \includegraphics[width=\textwidth]{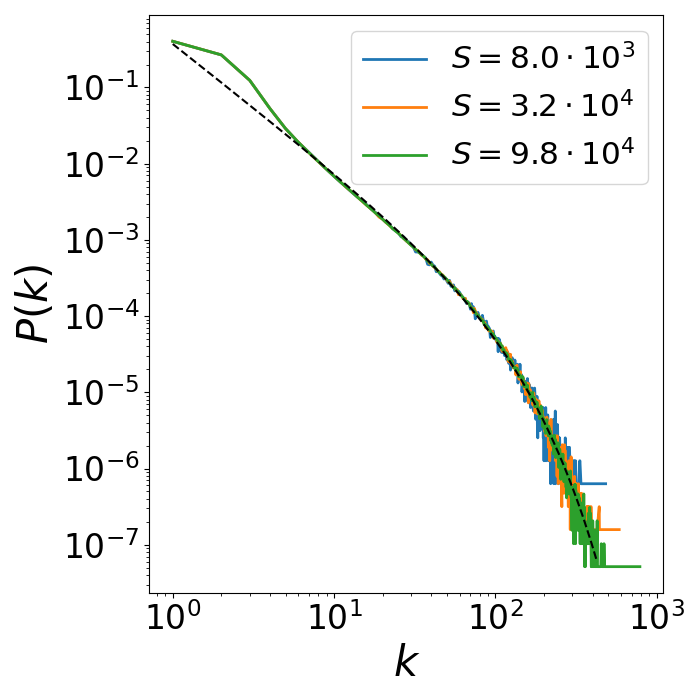}
        \includegraphics[width=\textwidth]{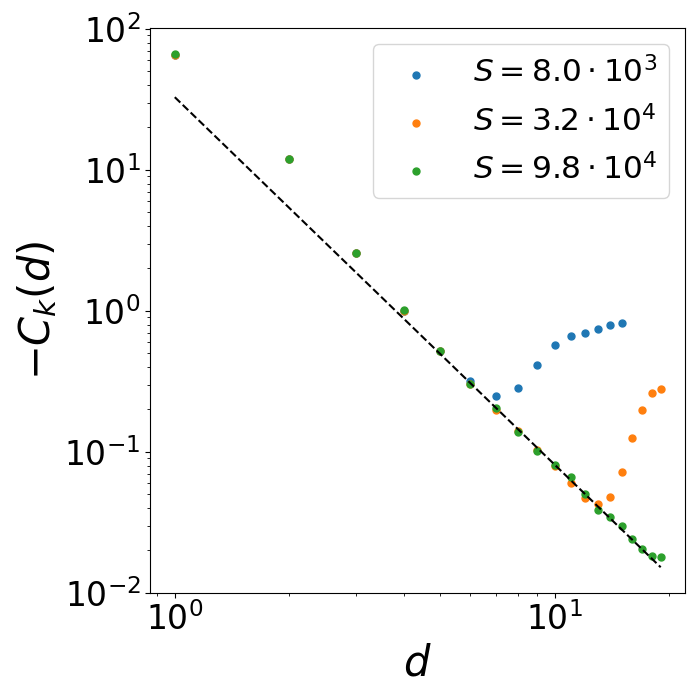}
        \includegraphics[width=\textwidth]{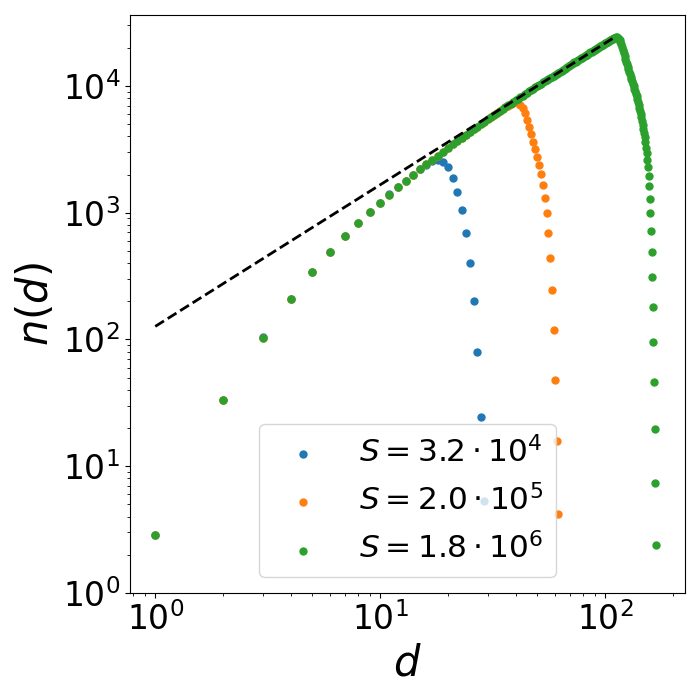}
        \caption{GKNS$_{2/5}$.}
    \end{subfigure}
    \begin{subfigure}[l]{.245\linewidth}
        \includegraphics[width=\textwidth]{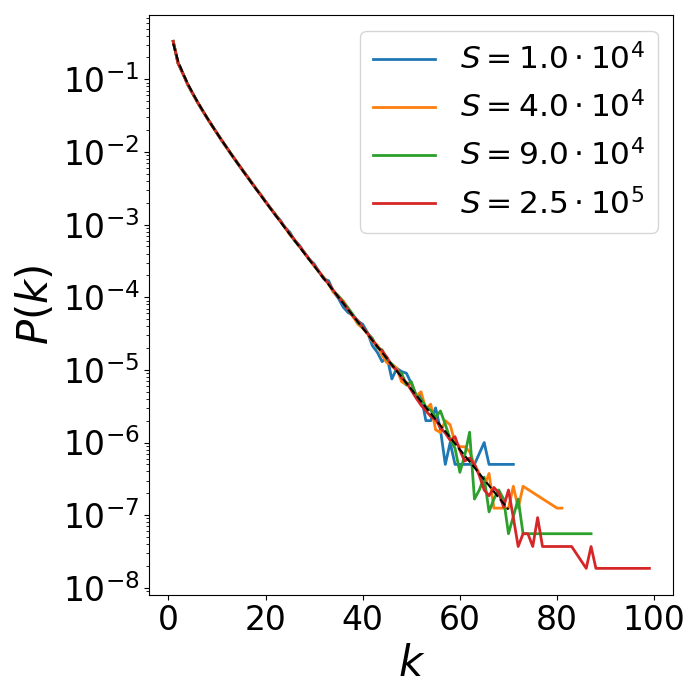}
        \includegraphics[width=\textwidth]{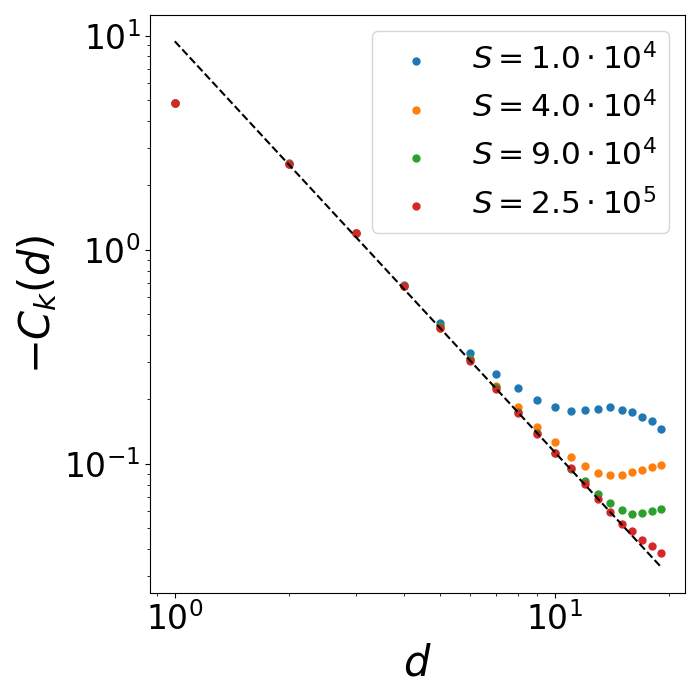}
        \includegraphics[width=\textwidth]{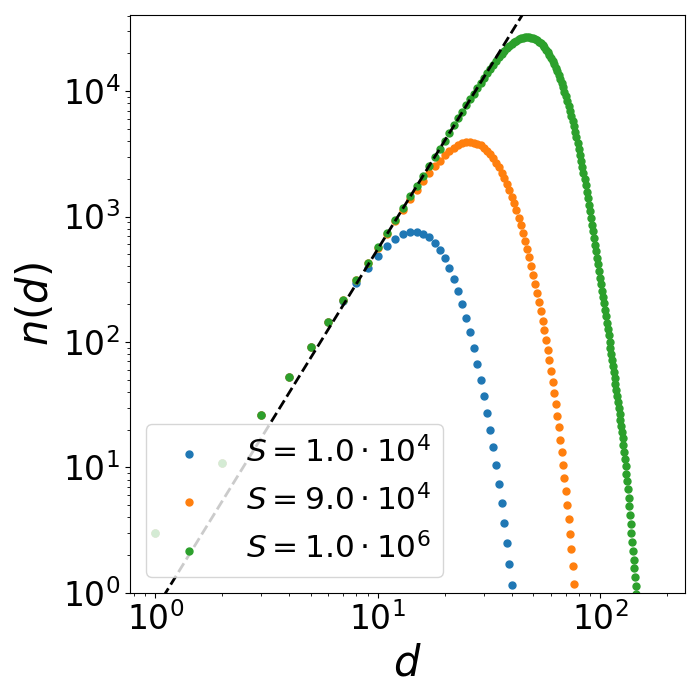}
        \caption{DQ.}
    \end{subfigure}
    \begin{subfigure}[l]{.245\linewidth}
        \includegraphics[width=\textwidth]{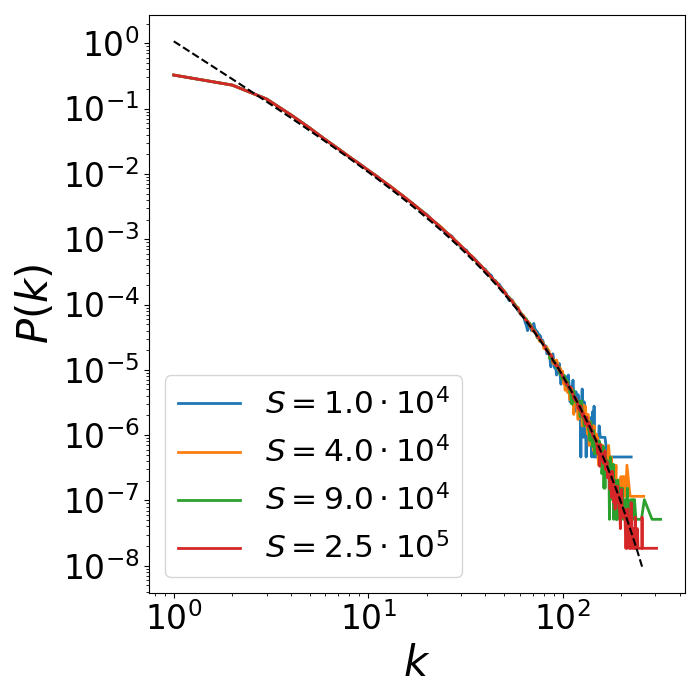}
        \includegraphics[width=\textwidth]{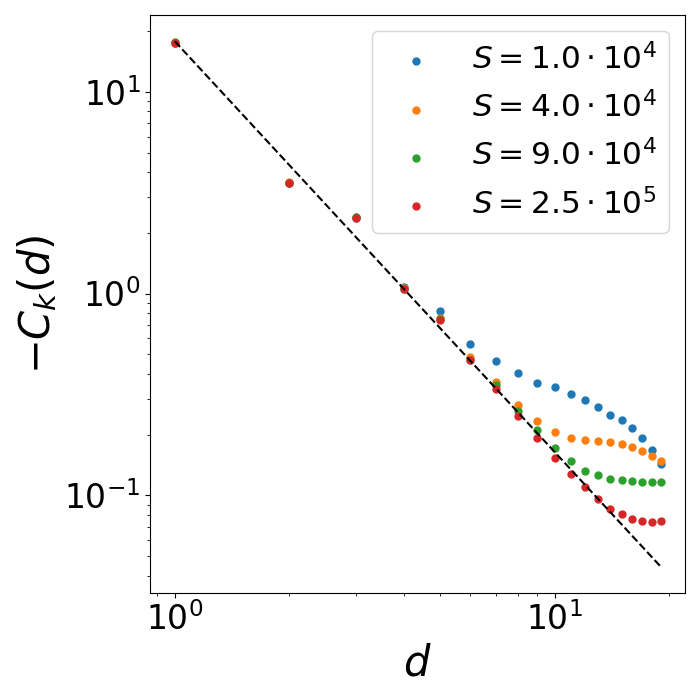}
        \includegraphics[width=\textwidth]{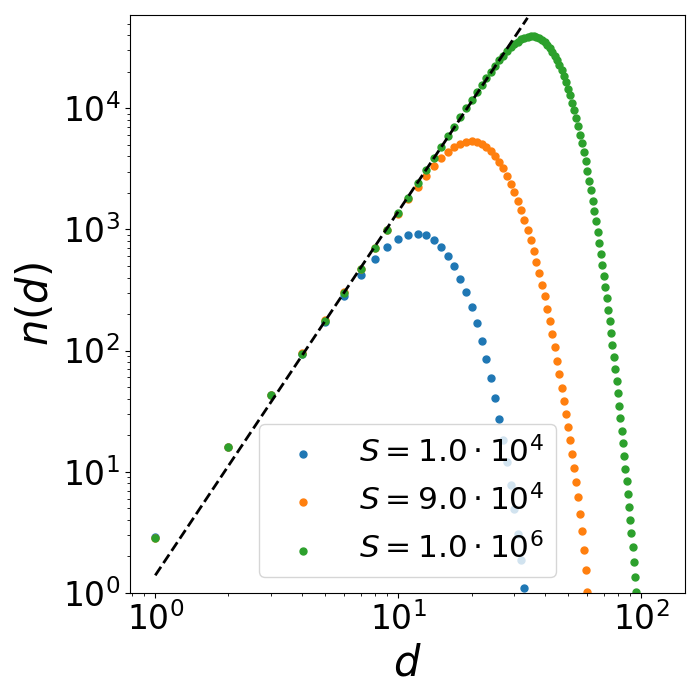}
        \caption{GQ.}
    \end{subfigure}
    \caption{The degree distribution $P(k)$ (top),
        the dependence of degree correlations $C_k(d)$ (middle)
        and the distance multiplicity $n(d)$ (bottom) on the distance $d$
        for GKNS, DQ and GQ surfaces
        with several values of surface area $S$ (in different colors).
        The approximating functions (black dashed lines)
        are of form $P(k) \propto k^{-\beta_k} \cdot e^{-k/k_0}$,
        $C_k(d) \propto d^{-\beta_c}$ and
        $n(d) \propto d^\varphi$ correspondingly,
        where the values $\beta_k$, $k_0$, $\beta_c$ and $\varphi$
        are free parameters determined by the best fit to the data
        (the precise values are given in \reft{scale_fits}).
        The $k$-axis scale in the $P(k)$ plot for DQ is linear,
        all the other scales are logarithmic.
        }
    \label{fig:scales}
\end{figure*}

\subsubsection{Degree distribution and correlations}

The degree distribution is an important characteristic of a random surface.
It has been shown to play a crucial role in the critical behavior of physical systems
\cite{Janke-2004, harris}.
Another relevant property is the degree-degree correlation
which is a key factor for understanding the clustering properties of
the surface \cite{deg_corrs-1, deg_corrs-2}.

The degree distribution $P(k)$ is defined as the probability
of a randomly chosen node to have degree $k$.
We define the degree correlations as
\begin{equation}
    C_k(d_0) = \inb<{\inb<{k_p k_q}>_{q, d(p,q)=d_0}}>_p - \inb<{k_p}>_p^2
\end{equation}
where $\inb<{\bullet}>_p$ is the average over all nodes $p$
and different realizations of the surface,
and the nested average is over nodes $q$ at distance $d(p,q)=d_0$
for a given node $p$. Notice that the introduced double averaging
is different from an averaging over all pairs of nodes at distance $d_0$.

$P(k)$ and $C_k(d)$ are studied numerically
for the discussed random surfaces with various values of surface area $S$
by analyzing over $200$ independent realizations for each case.
The data is portrayed in the upper two rows of \reff{scales}.

\begin{table}
\setlength{\tabcolsep}{7pt}
\begin{tabular}{r@{}r|cc|c|c}
    && $\beta_k$ & $k_0$ & $\beta_c$ & $\varphi$ \\ \hline
    GKNS & $_{1/3}$ & $1.64(5)$ & $20.9(4)$  & $2.47(4)$ & $1.06(4)$ \\
    GKNS & $_{2/5}$ & $1.66(3)$ & $76.2(16)$ & $2.61(4)$ & $1.12(8)$ \\
    GKNS & $_{4/9}$ & $1.65(2)$ & $431(6)$   & ---       & ---       \\
      DQ &          & $0.52(3)$ & $5.49(5)$  & $1.92(1)$ & $2.87(9)$ \\
      GQ &          & $1.88(4)$ & $31.4(6)$  & $2.04(6)$ & $3.00(3)$ \\
\end{tabular}
\caption{The exponents and other significant parameters
describing the GKNS, DQ and GQ surfaces:
degree distribution $P(k)\propto k^{\beta_k} \cdot e^{-k/k_0}$, 
degree correlations $C_k(d)\propto d^{-\beta_c}$ and
distance recurrence $n(d)\propto d^\varphi$}.
\label{tab:scale_fits}
\vspace{-1.cm}
\end{table}

The degree distributions for smaller degrees of GKNS and GQ surfaces
are in a power law, indicating a scale-free behavior \cite{Janke-2004},
however, for larger degrees they are exponentially suppressed.
The distribution for DQ surfaces exhibits exponential decay.
Generally the distributions can be well approximated by the form
$P(k) \propto k^{-\beta_k} \cdot e^{-k/k_0}$
with free parameters $\beta_k$ (the power-law exponent) and
$k_0$ (the exponential suppression scale),
and the normalization constant $\alpha_k$.
The corresponding values are reflected in \reft{scale_fits}.
Fairly big values of $k_0$ for GKNS and GQ indicate
the dominant power-law behavior,
whereas for DQ the exponential suppression is more significant.
It is important that the distributions don't depend on the surface area $S$.
Another important fact to note is the independence of the exponent $\beta_k\approx5/3$
on the value of $p$ on a GKNS$_p$ surfaces,
while $k_0$ increases and will reach infinity at criticality.

As for the degree correlations,
they are negative and decay with distance in a power law
for all the considered surfaces $C_k(d) \propto d^{-\beta_c}$.
The values of $\beta_c$ are shown in \reft{scale_fits}.
Unlike the degree distribution, $C_k(d)$ does depend on $S$,
however, the dependence is only through the finite-size effects at
distances comparable to the diameter (the largest distance) of the surface, $d \sim D$.
The correlations $C_k(d)$ otherwise coincide for different values of $S$ at smaller distances,
which indicates the existence of a well-defined thermodynamic limit.

It is also worth mentioning that due to computational limitations
the data for $C_k(d)$ is only available for $d \leq 20$ which
does not allow the detection of possible large-distance exponential suppression.

The frequently considered DT surface is known to have
an exponentially decaying degree distribution \cite{Ambjorn-1985, Ambjorn-book}.
In that sense, it is similar to DQ surfaces,
whereas the same comparison to GKNS and GQ surfaces shows
that the latter are essentially different.

\subsubsection{Hausdorff dimension}

To study surface scaling,
we introduce the notion of distance multiplicity $n(d)$ \cite{dist_mul}.
It shows the average number of nodes at a given distance $d$ from a node.
In the continuous regular extension, $n(d)$ is the surface area of a sphere of radius $d$.

It is reasonable to expect a power-law dependence
\begin{equation}
    n(d)\propto d^\varphi
\end{equation}
for large distances on an infinite system.
However, in a given realization there are short- and long-range deviations.
The source of short-range deviations is obviously the discrete nature of our surfaces,
as the scaling form works when the studied distances are closer to the continuum.
The long-range deviations appear because the considered systems are always compact.
To quantify this compactness, the notion of ``diameter" is introduced.
The diameter $D$ of a surface is the largest distance that exists on the surface
(the distance between the two nodes that are the farthest apart).
Thus, the scaling formula $n(d)\propto d^\varphi$ is valid for $1 \ll d \ll D$.

The ``volume" (number of nodes) $V(d)$ within distance $d$
is also expected to scale in a power law.
The corresponding exponent $\Delta_H$, for which $V(d) \propto d^{\Delta_H}$,
is the Hausdorff dimension of the surface.
As $V(d)=\sum_{d'=0}^d n(d')$,
switching to an integration for large values of $d'$ produces
\begin{equation}
    \Delta_H=\varphi+1 \squad.
\label{eq:hausdorff}
\end{equation}

The surface area $S$ and its diameter $D$ should also be linked through $\Delta_H$.
The corresponding relation is $D\propto S^{1/\Delta_H}$,
as $S$ is the ``volume" of the whole surface and $D$ is its ``linear" size.
For a regular square surface, $n_{\text{reg}}(d)=4d$,
so $\varphi_{\text{reg}}=1$ and $\Delta_{H, \text{reg}}=2$.
The scaling of the area with the diameter is $S_{\text{reg}}\propto D^2$.
Although all the surfaces considered are planar,
the values $\varphi=1$ and $\Delta_H=2$ do not necessarily hold.

The function $n(d)$ can be extracted by computing the distance multiplicity on surfaces.
The corresponding numerical data for the GKNS, DQ, and GQ surfaces
with several values of area $S$ are shown in the bottom row of \reff{scales}.
The data is collected over $100$ independent surface realizations for each case.
The determined values of the exponent $\varphi$ are given in \reft{scale_fits}.

Comparing again to the conventional DT surfaces,
which have Hausdorff dimension $\Delta_H^\text{DT}=4$ \cite{Ambjorn-book}
we see that DQ and GQ surfaces have a similar property
$\Delta_H^\text{DQ}\approx\Delta_H^\text{GQ}\approx 4$
whereas the GKNS surfaces are different again, with $\Delta_H^{\text{GKNS}}\approx 2$,
further emphasizing its uniqueness.

\subsubsection{Diameter}

\begin{figure}[t]
    \centering
    \includegraphics[width=.8\linewidth]{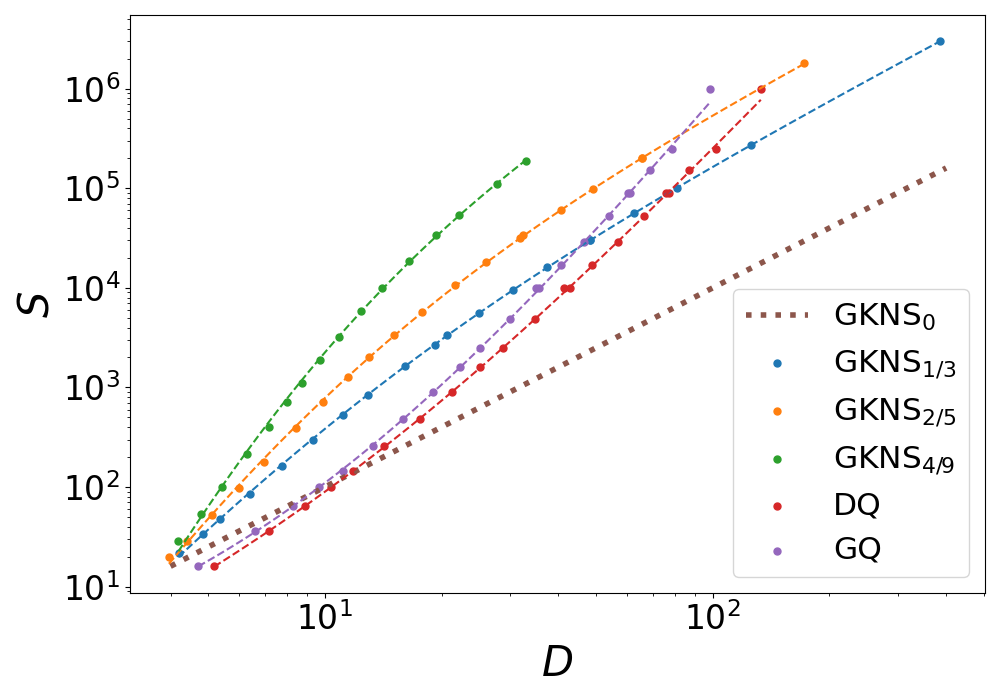}
    \caption{The scaling of surface area $S$ with diameter $D$
        for  GKNS, DQ, and GQ random surfaces (in different colors).
        The numerically obtained data (dots)
        and the fitting curves given by \refe{d_scaling} (dashed lines).
        The behavior for GKNS$_0$ is known exactly.}
    \label{fig:d_scaling}
\end{figure}

\begin{table}[t]
\setlength{\tabcolsep}{5pt}
\begin{tabular}{c|ccc|c|c}
    && GKSN$_p$ && DQ & GQ \vspace{-.12cm}\\ 
    $_p$ & $_{1/3}$ & $_{2/5}$ & $_{4/9}$ & $_-$ & $_-$ \\ \hline
    $\Delta_<$ & $4.26(10)$ & $5.9(3)$ & $8.9(6)$  & $1.7(5)$ & $0.7(9)$ \\
    $\Delta_>$ & $2.04(14)$ & $2.0(2)$ & $2.0(2)$ & $4.1(4)$ & $4.7(8)$ \\ \hline
    $D_0$ & $10.4(8)$ & $7.8(12)$ & $5.9(10)$ & $12(5)$ & $8(4)$ \\
\end{tabular}
\caption{The determined small and large scale dimensions $\Delta_<$ and $\Delta_>$,
    the crossover diameter $D_0$ 
    of GKNS, DQ and GQ surfaces.}
\label{tab:diams}
\end{table}

The connection between the surface area and the diameter
provides an alternative pathway for determining the Hausdorff dimension $\Delta_H$.
Numerical data for diameter $D$ are collected for several values of surface area $S$
for GKNS, DQ, and GQ random surfaces with $100-1000$ independent realizations each.
The data presented in \reff{d_scaling} reveals a crossover of $\Delta_H$
when transitioning from small to large systems.
We incorporate the aforementioned behavior into the scaling ansatz
\begin{equation}
    S \propto D^{\Delta_<} \cdot \inb({1 + D/D_0})^{\Delta_>-\Delta_<}
    \label{eq:d_scaling}
\end{equation}
where $D_0$ is the crossover scale between the two scaling regimes,
$\Delta_<$ and $\Delta_>$ are the scaling dimensions at small and large sizes, respectively.
The form ensures the $S \propto D^{\Delta_>}$ scaling at $D \gg D_0$ and
$S \propto D^{\Delta_<}$ at $D \ll D_0$.

The obtained values are shown in \reft{diams} and
are less accurate than those obtained via distance multiplicity calculations.
However, they confirm (don't contradict) the previously obtained values
$\Delta_H^\text{GKNS}\approx 2 \approx \Delta_>^\text{GKNS}$ and
$\Delta_H^\text{DQ/GQ}\approx 4 \approx \Delta_>^\text{DQ/GQ}$.

Another interesting feature is the increase of
$\Delta_<^{\text{GKNS}p}$ with increasing $p$.
Our numerical results indicate that
\begin{equation}
    \Delta_<^{\text{GKNS}_p}
    \propto (p-p_0)^{-\zeta}
    \squad,\squad \zeta=0.61(5) \squad.
\end{equation}
At the same time, the scaling regime crossover moves towards larger surface areas $S_0$,
despite the decrease of the corresponding crossover diameter $D_0$.
This behavior indicates that the short-distance geometry
becomes progressively more compact as $p$ approaches $p_0$,
whereas the asymptotic large-scale geometry remains two-dimensional.

\subsection{Comparison of GKNS, DQ and GQ to DT}

To place the properties of quadrangular random surfaces into context,
it is instructive to compare them with the established case of DT,
where many of their geometric and critical properties are known
from the extensive studies in statistical physics and two-dimensional quantum gravity.
\cite{Ambjorn-book,Kazakov-1985,AmbjornJurkiewiczWatabiki-1995,DurhuusJonssonWheater-2010,KawaiKawamotoMogamiWatabiki-1993,AmbjornJurkiewiczLoll-2005, AmbjornBuddWatabiki-2013,Watabiki-1995,Loll-2020}.

The determined local and global geometric characteristics
of quadrangular random surfaces provide complementary information for the comparison.
DQ is similar to DT in both its exponentially suppressed degree distribution and its
four-dimensional fractal scaling. GQ has the same global Hausdorff dimension but
a much broader degree distribution, indicating significant differences at
short and intermediate scales. GKNS surfaces differ from DT in both aspects:
their connectivity distribution approaches a scale-free form at criticality,
while their asymptotic Hausdorff dimension is close to two. These observations
support the conclusion that the GKNS critical geometry does not belong to the
ordinary two-dimensional Euclidean DT universality class.	

The obtained values of Hausdorff dimensions $\Delta_H$
reflected in \reft{scale_fits} (through \refe{hausdorff})
and \reft{diams} are consistently found to be
\begin{equation}
    \Delta_H^{\text{GKNS}}\approx 2
    \squad,\squad
    \Delta_H^{\text{DQ/GQ}}\approx 4 \squad.
\end{equation}
In two-dimensional Euclidean DT \cite{AmbjornJurkiewiczWatabiki-1995,DurhuusJonssonWheater-2010,KawaiKawamotoMogamiWatabiki-1993,AmbjornJurkiewiczLoll-2005}
describing pure quantum gravity, the typical geometry is strongly fractal and is characterized
by the intrinsic Hausdorff dimension
\begin{equation}
    \Delta_H^{\text{EDT}}=4 \squad.
\end{equation}
Consequently, the large-scale behavior of the DQ and GQ ensembles
$S\propto D^4$ allows them to be in the same geometric universality class, at least
in terms of their intrinsic Hausdorff dimension. This agreement is natural because
the quadrangulation and triangulation discretizations are generally expected to
produce the same continuum geometry when no additional constraint changes the
universality class.

By contrast, two-dimensional causal DT \cite{AmbjornBuddWatabiki-2013,Watabiki-1995,Loll-2020} are characterized by
\begin{equation}
    \Delta_H^{\text{CDT}}=2 \squad.
\end{equation}
The large-scale scaling of the GKNS ensembles $S\propto D^2$, is therefore closer to
that of causal DT than to ordinary Euclidean DT.
This comparison should nevertheless be understood at the level of
the Hausdorff dimension only. In particular, the condition defining the GKNS
ensemble is not equivalent to the causal foliation imposed in causal DT, and geometries
with the same Hausdorff dimension need not belong to the same universality class.
For example, smooth two-dimensional geometries and branched polymers may both
have $\Delta_H=2$, despite possessing substantially different local and global
structures. Additional observables, such as the spectral dimension, baby-universe
size distribution, susceptibility exponent, and connectivity correlations, are therefore
required to establish a stronger correspondence.

Scale-dependent effective Hausdorff dimensions and crossovers between distinct
geometric regimes are also known in generalized DT models,
including ensembles interpolating between ordinary and causal DT.
In such systems, different local and global Hausdorff dimensions
do not necessarily imply the existence of two distinct thermodynamic limits,
but they rather may describe a single continuum limit with a parameter-dependent
crossover scale.
Hence, the presented data do not singlehandedly demonstrate
the existence of two thermodynamic limits at criticality.
A second thermodynamic limit would require the crossover scale $S_0$
to diverge under an appropriate double-scaling procedure,
in which $S\to\infty$ and $p\to p_0$ are taken
simultaneously while a proper factor involving the ratio $S/S_0(p)$ is kept fixed.

Thus, for a fixed $p>p_0$, the most direct interpretation is that
$\Delta_<^{\text{GKNS}_p}$ represents a pre-asymptotic scale-dependent
effective dimension, while
\begin{equation}
    \Delta_H^{\text{GKNS}p}=\Delta_>^{\text{GKNS}p}\approx 2
\end{equation}
determines the thermodynamic limit.
Nevertheless, the apparent divergence of
$\Delta_<^{\text{GKNS}_p}$ and the growth of $S_0$ near $p_0$ suggest the
possible existence of a nontrivial double-scaling regime
which deserves a separate finite-size scaling analysis.

\section{Conclusion}

We have presented a systematic study of
random quadrangular surfaces and introduced a unified framework
encompassing three distinct ensembles:
the GKNS, dynamical quadrangulation (DQ), and general quadrangulation (GQ).
By establishing the relations between the elementary operations defining these ensembles,
we showed that they can be understood within a
common graph-theoretic description based on quadrangular lattices
and their dual representations.
This framework makes it possible to compare
different notions of randomness and provides a natural extension
of the dynamical triangulation (DT) paradigm to quadrangular geometries.

For the GKNS ensemble, we demonstrated its mapping
to two mutually constrained percolation processes
and determined the corresponding critical point and critical exponents.
The correlations between the two percolation sectors were shown
to modify the critical behavior relative to ordinary percolation,
leading to a distinct universality class.
For the DQ and GQ ensembles, we formulated the generation procedures
as ergodic Markov chains, analyzed their convergence properties,
and determined the scaling of the mixing and relaxation times with the surface area.
These results provide practical criteria for generating
statistically independent equilibrium realizations of quadrangular surfaces.

The statistical geometry of the three ensembles was characterized
through degree distributions, degree correlations, distance statistics
and Hausdorff dimensions.
We found that DQ surfaces resemble conventional DT surfaces,
exhibiting exponentially decaying degree distributions and
an asymptotic Hausdorff dimension $\Delta_H^{\text{DQ}} \approx 4$.
In contrast, GKNS and GQ surfaces develop broad degree distributions
with power-law behavior over a wide range of degrees.
Most remarkably, GKNS surfaces possess
an asymptotic Hausdorff dimension close to $\Delta_H^{\text{DQ}} \approx 2$.
Together with modified percolation criticality,
this demonstrates that GKNS quadrangular random surfaces
support universality classes that have no counterpart in DT.

These results suggest that quadrangulated random surfaces provide
families of random geometries that are a substantially different 
from the commonly considered triangular surfaces.
Since random triangulations furnish the discrete formulation
of two-dimensional quantum gravity and non-critical string theory,
the existence of new universality classes of quadrangular surfaces
raises the possibility of alternative continuum limits
based on quadrangulated geometries.
In particular, the GKNS universality class may provide
the geometric foundation for a new class of non-critical string theories,
extending the correspondence between discrete random surfaces
and continuum quantum gravity beyond the framework of triangulations.

\section*{Acknowledgments}
AS is thankful to Jan Ambj{\o}rn for useful discussions.
The research was supported by Armenian HESC
grant 24RL-1C024 (HT).

\bibliography{refs}

@BOOK{markov_chains,
  title     = "Markov chains and mixing times",
  author    = "Levin, David A and Peres, Yuval",
  publisher = "American Mathematical Society",
  edition   =  2,
  month     =  oct,
  year      =  2017,
  address   = "Providence, RI"
}

@article{gkns,
  title = {Geometrically disordered network models, quenched quantum gravity, and critical behavior at quantum Hall plateau transitions},
  author = {Gruzberg, I. A. and Kl\"umper, A. and Nuding, W. and Sedrakyan, A.},
  journal = {Phys. Rev. B},
  volume = {95},
  issue = {12},
  pages = {125414},
  numpages = {12},
  year = {2017},
  month = {Mar},
  publisher = {American Physical Society},
  doi = {10.1103/PhysRevB.95.125414}
}

@article{conti,
  title = {Geometry of random potentials: Induction of two-dimensional gravity in quantum Hall plateau transitions},
  author = {Conti, Riccardo and Topchyan, Hrant and Tateo, Roberto and Sedrakyan, Ara},
  journal = {Phys. Rev. B},
  volume = {103},
  issue = {4},
  pages = {L041302},
  numpages = {5},
  year = {2021},
  month = {Jan},
  publisher = {American Physical Society},
  doi = {10.1103/PhysRevB.103.L041302},
  nurl = {https://link.aps.org/doi/10.1103/PhysRevB.103.L041302}
}

@article{s-matrix,
  title = {Integer quantum Hall transition: An $S$-matrix approach to random networks},
  author = {Topchyan, H. and Gruzberg, I. A. and Nuding, W. and Kl\"umper, A. and Sedrakyan, A.},
  journal = {Phys. Rev. B},
  volume = {110},
  issue = {8},
  pages = {L081112},
  numpages = {6},
  year = {2024},
  month = {Aug},
  publisher = {American Physical Society},
  doi = {10.1103/PhysRevB.110.L081112},
  nurl = {https://link.aps.org/doi/10.1103/PhysRevB.110.L081112}
}

@article{barabasi,
  title = {Statistical mechanics of complex networks},
  author = {Albert, R\'eka and Barab\'asi, Albert-L\'aszl\'o},
  journal = {Rev. Mod. Phys.},
  volume = {74},
  issue = {1},
  pages = {47--97},
  numpages = {0},
  year = {2002},
  month = {Jan},
  publisher = {American Physical Society},
  doi = {10.1103/RevModPhys.74.47},
}

@inproceedings{netlsd,
author = {Tsitsulin, Anton and Mottin, Davide and Karras, Panagiotis and Bronstein, Alexander and M\"{u}ller, Emmanuel},
title = {NetLSD: Hearing the Shape of a Graph},
year = {2018},
isbn = {9781450355520},
publisher = {Association for Computing Machinery},
nurl = {https://doi.org/10.1145/3219819.3219991},
doi = {10.1145/3219819.3219991},
booktitle = {Proceedings of the 24th ACM SIGKDD International Conference on Knowledge Discovery \& Data Mining},
pages = {2347–2356},
numpages = {10},
series = {KDD '18}
}

@article{slq,
author = {Ubaru, Shashanka and Chen, Jie and Saad, Yousef},
title = {Fast Estimation of $tr(f(A))$ via Stochastic Lanczos Quadrature},
journal = {SIAM Journal on Matrix Analysis and Applications},
volume = {38},
number = {4},
pages = {1075-1099},
year = {2017},
doi = {10.1137/16M1104974},
nURL = {https://doi.org/10.1137/16M1104974}
}

@article{autocorr,
  title = {Analysis of autocorrelation times in neural Markov chain Monte Carlo simulations},
  author = {Bia\l{}as, Piotr and Korcyl, Piotr and Stebel, Tomasz},
  journal = {Phys. Rev. E},
  volume = {107},
  issue = {1},
  pages = {015303},
  numpages = {18},
  year = {2023},
  month = {Jan},
  publisher = {American Physical Society},
  doi = {10.1103/PhysRevE.107.015303},
  nurl = {https://link.aps.org/doi/10.1103/PhysRevE.107.015303}
}

@article{Janke-2004,
  title = {Harris-Luck criterion for random lattices},
  author = {Janke, Wolfhard and Weigel, Martin},
  journal = {Phys. Rev. B},
  volume = {69},
  issue = {14},
  pages = {144208},
  numpages = {12},
  year = {2004},
  month = {Apr},
  publisher = {American Physical Society},
  doi = {10.1103/PhysRevB.69.144208},
  nurl = {https://link.aps.org/doi/10.1103/PhysRevB.69.144208}
}

@article{harris,
  title = {Harris-Luck criterion in the plateau transition of the integer quantum Hall effect},
  author = {Topchyan, H. and Nuding, W. and Kl\"umper, A. and Sedrakyan, A.},
  journal = {Phys. Rev. B},
  volume = {111},
  issue = {10},
  pages = {L100201},
  numpages = {6},
  year = {2025},
  month = {Mar},
  publisher = {American Physical Society},
  doi = {10.1103/PhysRevB.111.L100201},
  nurl = {https://link.aps.org/doi/10.1103/PhysRevB.111.L100201}
}

@article{David-1985,
  title = {A model of random surfaces with non-trivial critical behaviour},
  volume = {257},
  ISSN = {0550-3213},
  url = {http://dx.doi.org/10.1016/0550-3213(85)90363-3},
  DOI = {10.1016/0550-3213(85)90363-3},
  journal = {Nuclear Physics B},
  publisher = {Elsevier BV},
  author = {David,  F.},
  year = {1985},
  month = jan,
  pages = {543–576}
}

@article{Ambjorn-1985,
title = {Diseases of triangulated random surface models, and possible cures},
journal = {Nuclear Physics B},
volume = {257},
pages = {433-449},
year = {1985},
issn = {0550-3213},
doi = {https://doi.org/10.1016/0550-3213(85)90356-6},
author = {J. Ambjørn and B. Durhuus and J. Fröhlich},
}

@article{Kazakov-1985,
  title = {Bilocal regularization of models of random surfaces},
  volume = {150},
  ISSN = {0370-2693},
  url = {http://dx.doi.org/10.1016/0370-2693(85)91011-1},
  DOI = {10.1016/0370-2693(85)91011-1},
  number = {4},
  journal = {Physics Letters B},
  publisher = {Elsevier BV},
  author = {Kazakov,  V.A.},
  year = {1985},
  month = jan,
  pages = {282–284}
}

@article{Polyakov-1981,
  author = {Polyakov, Alexander M.},
  title = {Quantum Geometry of Bosonic Strings},
  journal = {Physics Letters B},
  volume = {103},
  number = {3},
  pages = {207--210},
  year = {1981}
}

@book{Polyakov-book,
  author    = {Polyakov, Alexander M.},
  title     = {Gauge Fields and Strings},
  year      = {1987},
  publisher = {Harwood Academic Publishers}
}

@book{Ambjorn-book,
  author    = {Ambj{\o}rn, Jan and Durhuus, Bergfinnur and J{\'o}nsson, Thordur},
  title     = {Quantum Geometry: A Statistical Field Theory Approach},
  publisher = {Cambridge University Press},
  year      = {1997},
  address   = {Cambridge},
  isbn      = {978-0521461678}
}

@article{KPZ-1988,
  author  = {Knizhnik, V. G. and Polyakov, A. M. and Zamolodchikov, A. B.},
  title   = {Fractal Structure of 2D Quantum Gravity},
  journal = {Modern Physics Letters A},
  volume  = {3},
  number  = {8},
  pages   = {819--826},
  year    = {1988},
  doi     = {10.1142/S0217732388000989}
}

@article{Distler-1989,
  author  = {Distler, Jacques and Kawai, Hikaru},
  title   = {Conformal Field Theory and 2D Quantum Gravity},
  journal = {Nuclear Physics B},
  volume  = {321},
  number  = {2},
  pages   = {509--527},
  year    = {1989},
  doi     = {10.1016/0550-3213(89)90354-4}
}

@article{David-1988,
  author  = {David, Francois},
  title   = {Conformal Field Theories Coupled to 2D Gravity in the Conformal Gauge},
  journal = {Modern Physics Letters A},
  volume  = {3},
  number  = {17},
  pages   = {1651--1656},
  year    = {1988},
  doi     = {10.1142/S0217732388001971}
}

@article{Duplantier-1988,
  author = {Duplantier, Bertrand and Kostov, Ivan K.},
  title = {Conformal spectra of polymers on random surfaces},
  journal = {Physical Review Letters},
  volume = {61},
  number = {13},
  pages = {1430--1433},
  year = {1988},
  doi = {10.1103/PhysRevLett.61.1430}
}

@article{Kostov-1989,
  title = {Random surfaces,  solvable lattice models and discrete quantum gravity in two dimensions},
  volume = {10},
  ISSN = {0920-5632},
  url = {http://dx.doi.org/10.1016/0920-5632(89)90069-8},
  DOI = {10.1016/0920-5632(89)90069-8},
  number = {1},
  journal = {Nuclear Physics B - Proceedings Supplements},
  publisher = {Elsevier BV},
  author = {Kostov,  I.K.},
  year = {1989},
  month = Jul,
  pages = {295–322}
}

@article{Duplantier-1989,
  author = {Duplantier, Bertrand},
  title = {Statistical mechanics of polymer networks of any topology},
  journal = {Journal of Statistical Physics},
  volume = {54},
  number = {3-4},
  pages = {581--680},
  year = {1989},
  doi = {10.1007/BF01019738}
}

@article{Wegner-1971,
  author = {Wegner, Franz J.},
  title = {Duality in generalized Ising models and phase transitions without local order parameters},
  journal = {Journal of Mathematical Physics},
  volume = {12},
  number = {10},
  pages = {2259--2272},
  year = {1971},
  doi = {10.1063/1.1665530}
}

@article{Wilson-1974,
  author = {Wilson, Kenneth G. and Kogut, John},
  title = {The renormalization group and the epsilon expansion},
  journal = {Physics Reports},
  volume = {12},
  number = {2},
  pages = {75--199},
  year = {1974},
  doi = {10.1016/0370-1573(74)90023-4}
}

@article{deg_corrs-1,
  title = {Mixing patterns in networks},
  author = {Newman, M. E. J.},
  journal = {Phys. Rev. E},
  volume = {67},
  issue = {2},
  pages = {026126},
  numpages = {13},
  year = {2003},
  month = {Feb},
  publisher = {American Physical Society},
  doi = {10.1103/PhysRevE.67.026126},
  url = {https://link.aps.org/doi/10.1103/PhysRevE.67.026126}
}

@article{deg_corrs-2,
title = {Higher order assortativity in complex networks},
journal = {European Journal of Operational Research},
volume = {262},
number = {2},
pages = {708-719},
year = {2017},
issn = {0377-2217},
doi = {https://doi.org/10.1016/j.ejor.2017.04.028},
url = {https://www.sciencedirect.com/science/article/pii/S0377221717303612},
author = {Alberto Arcagni and Rosanna Grassi and Silvana Stefani and Anna Torriero},
}

@article{dist_mul,
    title={Spherical averages of Fourier transforms of measures with finite energy; dimensions of intersections and distance sets},
    volume={34},
    doi={10.1112/S0025579300013462},
    number={2},
    journal={Mathematika},
    author={Mattila, Pertti},
    year={1987},
    pages={207–228}
}

@book{percolation,
  title = {Introduction To Percolation Theory},
  ISBN = {9781482272376},
  url = {http://dx.doi.org/10.1201/9781315274386},
  DOI = {10.1201/9781315274386},
  publisher = {Taylor \& Francis},
  author = {Stauffer,  Dietrich and Aharony,  Ammon},
  year = {2018},
  month = Dec 
}

@book{CG-tris,
  title = {Polygon Mesh Processing},
  ISBN = {9781439865316},
  url = {http://dx.doi.org/10.1201/b10688},
  DOI = {10.1201/b10688},
  publisher = {A K Peters/CRC Press},
  author = {Botsch,  Mario and Kobbelt,  Leif and Pauly,  Mark and Alliez,  Pierre and Levy,  Bruno},
  year = {2010},
  month = Oct 
}

@article{AmbjornJurkiewiczWatabiki-1995,
  author  = {J. Ambj{\o}rn and J. Jurkiewicz and Y. Watabiki},
  title   = {On the Fractal Structure of Two-Dimensional Quantum Gravity},
  journal = {Nuclear Physics B},
  volume  = {454},
  pages   = {313--342},
  year    = {1995},
  doi     = {10.1016/0550-3213(95)00399-6},
  eprint  = {hep-lat/9507014},
  archivePrefix = {arXiv}
}

@article{DurhuusJonssonWheater-2010,
  author  = {B. Durhuus and T. Jonsson and J. F. Wheater},
  title   = {The Spectral Dimension of Generic Trees},
  journal = {Journal of Statistical Physics},
  volume  = {139},
  pages   = {859--881},
  year    = {2010},
  doi     = {10.1007/s10955-010-9979-5},
  eprint  = {0908.3643},
  archivePrefix = {arXiv},
  primaryClass = {hep-th}
}

@article{AmbjornBuddWatabiki-2013,
  author  = {J. Ambj{\o}rn and T. G. Budd and Y. Watabiki},
  title   = {The Universe from Scratch},
  journal = {Physics Letters B},
  volume  = {728},
  pages   = {58--63},
  year    = {2014},
  doi     = {10.1016/j.physletb.2013.11.043},
  eprint  = {1302.1763},
  archivePrefix = {arXiv},
  primaryClass = {hep-th}
}

@article{Watabiki-1995,
  author  = {Y. Watabiki},
  title   = {Construction of Non-Critical String Field Theory by Transfer Matrix Formalism in Dynamical Triangulation},
  journal = {Nuclear Physics B},
  volume  = {441},
  pages   = {119--163},
  year    = {1995},
  doi     = {10.1016/0550-3213(95)00091-P},
  eprint  = {hep-th/9401096},
  archivePrefix = {arXiv}
}

@article{KawaiKawamotoMogamiWatabiki-1993,
  author  = {H. Kawai and N. Kawamoto and T. Mogami and Y. Watabiki},
  title   = {Transfer Matrix Formalism for Two-Dimensional Quantum Gravity and Fractal Structures of Space-Time},
  journal = {Physics Letters B},
  volume  = {306},
  pages   = {19--26},
  year    = {1993},
  doi     = {10.1016/0370-2693(93)90033-2},
  eprint  = {hep-th/9302133},
  archivePrefix = {arXiv}
}

@article{AmbjornJurkiewiczLoll-2005,
  author  = {J. Ambj{\o}rn and J. Jurkiewicz and R. Loll},
  title   = {Reconstructing the Universe},
  journal = {Physical Review D},
  volume  = {72},
  pages   = {064014},
  year    = {2005},
  doi     = {10.1103/PhysRevD.72.064014},
  eprint  = {hep-th/0505154},
  archivePrefix = {arXiv}
}

@article{Loll-2020,
  author  = {R. Loll},
  title   = {Quantum Gravity from Causal Dynamical Triangulations: A Review},
  journal = {Classical and Quantum Gravity},
  volume  = {37},
  number  = {1},
  pages   = {013002},
  year    = {2020},
  doi     = {10.1088/1361-6382/ab57c7},
  eprint  = {1905.08669},
  archivePrefix = {arXiv},
  primaryClass = {hep-th}
}

@misc{dataset,
  doi = {10.5281/ZENODO.23604122},
  url = {https://zenodo.org/doi/10.5281/zenodo.23604122},
  author = {Topchyan,  Hrant and Sedrakyan,  Ara},
  title = {Statistical properties of quadrangular surfaces},
  publisher = {Zenodo},
  year = {2026},
  copyright = {Creative Commons Attribution 4.0 International}
}

@article{arboreal-1,
  title = {Loopless percolation clusters},
  volume = {39},
  ISSN = {0556-2791},
  url = {http://dx.doi.org/10.1103/PhysRevA.39.5470},
  DOI = {10.1103/physreva.39.5470},
  number = {10},
  journal = {Physical Review A},
  publisher = {American Physical Society (APS)},
  author = {Tzschichholz,  Frank and Bunde,  Armin and Havlin,  Shlomo},
  year = {1989},
  month = May,
  pages = {5470–5473}
}

@article{arboreal-2,
  title = {Random Spanning Forests and Hyperbolic Symmetry},
  volume = {381},
  ISSN = {1432-0916},
  url = {http://dx.doi.org/10.1007/s00220-020-03921-y},
  DOI = {10.1007/s00220-020-03921-y},
  number = {3},
  journal = {Communications in Mathematical Physics},
  publisher = {Springer Science and Business Media LLC},
  author = {Bauerschmidt,  Roland and Crawford,  Nicholas and Helmuth,  Tyler and Swan,  Andrew},
  year = {2020},
  month = Dec,
  pages = {1223–1261}
}

@article{chal-codd,
  title = {Percolation,  quantum tunnelling and the integer Hall effect},
  volume = {21},
  ISSN = {0022-3719},
  url = {http://dx.doi.org/10.1088/0022-3719/21/14/008},
  DOI = {10.1088/0022-3719/21/14/008},
  number = {14},
  journal = {Journal of Physics C: Solid State Physics},
  publisher = {IOP Publishing},
  author = {Chalker,  J T and Coddington,  P D},
  year = {1988},
  month = May,
  pages = {2665–2679}
}

\end{document}